\documentclass[aps,twocolumn,showpacs,preprintnumbers,floatfix]{revtex4}
\usepackage{graphicx}
\usepackage{dcolumn}
\usepackage{bm}
\begin{document}
\preprint{UK/05-09}
\preprint{BIHEP-TH-05-15}

\title{Glueball Spectrum and Matrix Elements on Anisotropic Lattices}

\author{
        Y. Chen${}^{a,b}$,
        A. Alexandru${}^{a}$, S.J. Dong${}^{a}$, T. Draper${}^a$,
        I. Horv\'{a}th${}^a$,
        F.X. Lee${}^{c,d}$,
        K.F. Liu${}^{a}$,\\
        N. Mathur${}^{a,d}$,
        C. Morningstar${}^e$,
        M. Peardon${}^{f}$,
        S. Tamhankar${}^a$,
        B.L. Young${}^g$, and
        J.B. Zhang${}^h$
       }
\affiliation{
  ${}^a$Department of Physics \& Astronomy, University of Kentucky, Lexington, KY 40506, USA\\
  ${}^b$Institute of High Energy Physics,
        Chinese Academy of Sciences, Beijing 100039, P.R. China\\
  ${}^c$Center for Nuclear Studies, Department of Physics,
        George Washington University, Washington, DC 20052 USA\\
  ${}^d$Jefferson Lab, 12000 Jefferson Avenue, Newport News, VA 23606, USA\\
  ${}^e$Department of Physics, Carnegie Mellon University, Pittsburgh,
PA 15213, USA\\
  ${}^f$School of Mathematics, Trinity College, Dublin, Dublin 2, Ireland\\
  ${}^g$Department of Physics and Astronomy,
        Iowa State University, Ames, Iowa 50011, USA\\
  ${}^h$CSSM and Department of Physics,
        University of Adelaide, Adelaide, SA 5005, Australia
           }
\begin{abstract}
{The glueball-to-vacuum matrix elements of local gluonic operators
in scalar, tensor, and pseudoscalar channels are investigated numerically
on several anisotropic lattices with the spatial lattice spacing ranging
from 0.1fm - 0.2fm. These matrix elements are needed to predict the
glueball branching ratios in $J/\psi$ radiative decays which will help
identify the glueball states in experiments.
Two types of improved local gluonic operators are
constructed for a self-consistent check and the finite volume effects
are studied. We find that lattice spacing dependence of our results is
very weak and the continuum limits are reliably extrapolated, as a
result of improvement of the lattice gauge action and local operators.
We also give updated glueball masses with various quantum numbers.}
\end{abstract}
\maketitle

\section{Introduction}
Glueballs, predicted by QCD, are so exotic from the point of view
of naive quark model that their existence will be a direct support
of QCD. However, experimental efforts in searching for glueballs are
confronted with the difficulty of identifying glueballs
unambiguously, even though there are several candidate glueball
resonances, such as $f_0(1370)$, $f_0(1500)$, $f_0(1710)$, and
$f_J(2220)$, etc.. The key problem is that there is little knowledge of 
the
nature of glueballs and confined QCD vacuum, which requires
reliable nonperturbative methods to be implemented. The numerical
study of lattice QCD, which starts from the first principles,
has been playing an important role in this hot field in the last
twenty years, and extensive numerical studies have been carried
out to calculate the glueball spectrum~\cite{berg,old1,old2,old3}.
These studies give the result that the masses of the lowest-lying
glueballs range from $1\,{\rm GeV}$ to $3\,{\rm GeV}$, and suggest that 
the $J/\psi$
radiative decays be an ideal hunting ground for glueballs.
However, apart from the mass spectrum, more characteristics are
desired in order to determine glueballs in the final states of
$J/\psi$ radiative decays, one of which is the partial widths of
$J/\psi$ decaying into glueballs. The first step to estimate these
partial widths is to calculate the vacuum-to-glueball transition
matrix elements(TME) of local gluonic operators, which are
nonperturbative quantities and can be investigated by the
numerical calculation of lattice gauge theory.
\par
The techniques of lattice calculations in the glueball sector have
been substantially improved in the past decade. It is known that
very large statistics are necessary in order for the correlation
functions of gluonic operators to be measured precisely by
Monte-Carlo simulation. This prohibits the lattice size being too
large. On the other hand, because of the large masses of
glueballs, the lattice spacing (at least in the temporal
direction) has to be small enough so that reliable signals can be
measured before they are undermined by statistical fluctuations.
This dilemma is circumvented by using anisotropic lattices, which are
spatially coarse and temporally fine. The potentially large
lattice artifacts owing to the spatially coarse lattice can be
suppressed by the implementation of improved lattice gauge
actions. These techniques has been verified to be efficient in the
calculations of glueball spectrum~\cite{old3}, and are adopted as the
basic formalism of this work.
\par
The glueball matrix elements computed in this work are of the form
$\langle 0|g^2 {\rm Tr} G_{\mu\nu}G_{\rho\sigma}(x)|G\rangle$, where
$|G\rangle$ refers to the glueball state with specific quantum
number $J^{PC}=0^{++}$, $2^{++}$, or $0^{-+}$, $G_{\mu\nu}(x)$ is
the gauge field strength, and $g$ the gauge coupling. The lattice
version of the gluonic operators
$g^2{\rm Tr}G_{\mu\nu}G_{\rho\sigma}(x)$ are constructed by the 
smallest
Wilson loops on the lattice. To reduce lattice artifacts,  the
lattice version of each local gluonic operator is improved by
eliminating the  $O(a_s^2)$ ($a_s$ here is the spatial lattice
spacing), while the glueball states $|G\rangle$ are obtained by
smeared gluonic operators. In order to get a reliable continuum
extrapolation,  five independent calculations are carried out on 
lattices with spatial lattice spacings ranging from 0.09 fm to
0.22 fm.  As by-products, we also calculate the masses of the
lowest-lying stationary states in all of the symmetry channels
allowed on a cubic lattice at each lattice spacing. To study the
systematic error from finite volume, two extra independent studies
are performed at $\beta = 2.4$ with the same input parameters but
different lattice size. Note that the local operators mentioned
above are all bare operators, they need to be renormalized to give
the physical matrix elements. In this work, the renormalization
constants of the scalar and tensor gluonic operators at the
largest lattice spacing are extracted by the calculations of gluonic
three-point functions on a much larger statistical sample (100,000
measurements). The pseudoscalar operator renormalization is
determined through the calculation of the topological
susceptibility. Finally, we give the nonperturbatively calculated
and phenomenologically significant matrix elements.
\par
This paper is organized as follows. In Section II, we give a
detailed description of the construction of lattice local gluonic
operators. Two types of lattice realization in each channel are
defined, including the improvement schemes. The details of the
computation, including the generation of the gauge configurations,
the construction of the smeared glueball correlators, the
extraction of glueball masses from the correlation functions, and
the determination of the lattice spacing, are described in Sec.
III. In Sec. IV, we estimate the finite volume errors and
some other systematical errors. The removal of lattice artifacts
due to the finite lattice spacing, including the extrapolations to
the continuum limit, is described in Sec. V. The calculation of
three point function and the calculation of renormalization
constants are described in Sec. VI. Sec. VII gives the conclusion and
some discussion.

\section{Local Gluonic Operators and Matrix Elements}
In the continuum theory, the lowest-dimensional gauge invariant gluonic
operators are of the form $g^2TrG_{\mu\nu}G_{\rho\sigma}$, of which the most
commonly studied ones are the scalar $S(x)$ (QCD trace anomaly), the
pseudoscalar $P(x)$ (topological charge density), and the tensor
$\Theta_{\mu\nu}(x)$ (energy-momentum density). They all have
dimension four and are all positive under charge
conjugation. The explicit forms of them are
\begin{eqnarray}
\label{lo}
S(x)&=& g^2{\rm Tr}G_{\mu\nu}(x)G_{\mu\nu}(x),\nonumber\\
P(x)&=& g^2 \epsilon_{\mu\nu\rho\sigma}{\rm Tr}
G_{\mu\nu}(x)G_{\rho\sigma}(x),\nonumber\\
\Theta_{\mu\nu}(x)&=& 2g^2 {\rm Tr}(G_{\mu\alpha}(x)G_{\alpha\nu}(x)
-\frac{1}{4}\delta_{\mu\nu}G^2(x)).
\end{eqnarray}

If we introduce the chromo-electric and chromo-magnetic fields,
\begin{equation}
\label{def0}
E_i = G_{i0}~~~{\rm and}~~~B_i = -\frac{1}{2}\epsilon_{ijk}G_{jk},
\end{equation}
the Lorentz scalar $S(x)$ and pseudoscalar $P(x)$ can be
expressed explicitly by these operators,
\begin{eqnarray}
\label{def1}
S(x) &=& 2 g^2 {\rm Tr}({\bf B}(x)^2+{\bf E}(x)^2),\nonumber\\
P(x) &=& 8 g^2 {\rm Tr}({\bf E}(x) \cdot {\bf B}(x)),\nonumber\\
\Theta_{ij}(x) &=& 2 g^2
{\rm Tr}(\theta_{ij}^B(x)-\theta_{ij}^E(x) ),
\end{eqnarray}
where $\theta^E$ and $\theta^B$ are the traceless, symmetric
chromo-electric and magnetic tensors,
\begin{eqnarray}
\label{def1.5}
\theta_{ij}^E &=& {\rm Tr} E_i E_j-\frac{1}{3}\delta_{ij}{\rm Tr}{\bf
E}^2\nonumber\\
\theta_{ij}^B &=& {\rm Tr} B_i B_j-\frac{1}{3}\delta_{ij}{\rm Tr}{\bf 
B}^2.
\end{eqnarray}
The scalar operators ${\rm Tr}{\bf B}^2(x)$ and ${\rm Tr}{\bf E}^2(x)$, 
the 
pseudoscalar
${\rm Tr} {\bf E}(x)\cdot {\bf B}(x)$, and the tensors $\theta^E$, 
$\theta^B$,
are all irreducible representations of the three-dimensional rotation
group $SO(3)$.
\par
Denoting the normalized scalar, pseudoscalar, and the tensor glueball
states as $|S\rangle$, $|P\rangle$, and $|T_{ij}\rangle$, respectively,
the non-zero glueball-to-vacuum matrix elements for their annihilation 
at rest by these operators are
\begin{eqnarray}
(2\pi)^3\delta^3({\bf 0})f_{(S,B)}&=&g^2\langle 0|\int d^3 x {\rm Tr} 
{\bf
B}^2(x)|S\rangle\nonumber\\
(2\pi)^3\delta^3({\bf 0})f_{(S,E)}&=&g^2\langle 0|\int d^3 x {\rm Tr} 
{\bf
E}^2(x)|S\rangle\nonumber\\
(2\pi)^3\delta^3({\bf 0})f_{(T,B)}&=&g^2\langle 0|\int d^3 x {\rm Tr}
\theta_{ij}^B(x)|T_{ij}\rangle\nonumber\\
(2\pi)^3\delta^3({\bf 0})f_{(T,E)}&=&g^2\langle 0|\int d^3 x {\rm Tr}
\theta_{ij}^E(x)|T_{ij}\rangle\nonumber\\
(2\pi)^3\delta^3({\bf 0})f_{(PS)}&=&g^2\langle 0|\int d^3 x {\rm Tr} 
{\bf
E}(x)\cdot {\bf B}(x)|P\rangle,
\end{eqnarray}
where no implicit summation is applied. It is straightforward to
reproduce the matrix elements of the scalar $S(x)$ and pseudoscalar
operator $P(x)$ by these quantities, but for the Lorentz tensor
$\Theta_{\mu\nu}$, the situation is more complicated. First of all, any hadron
state $|h\rangle$ is an eigenstate of $\Theta_{0\mu}$, so that the
matrix elements $\langle 0|\Theta_{0\mu}(x)|h\rangle$ are zero.
Together with the traceless property 
($\sum\limits_{\mu}\Theta_{\mu\mu}=0$), the above condition
 implies that $\langle 0|\sum\limits_i \Theta_{ii}(x)|h\rangle$ is zero. 
Therefore, there
are only five
linearly independent no-zero matrix elements of $\Theta_{\mu\nu}$,
which can be decomposed into the color-magnetic part and the
color-electric part,
\begin{equation}
\label{def2}
\langle 0|\Theta_{ij}(x)|T_{ij}\rangle = 2\langle
0|\theta_{ij}^B(x)-\theta_{ij}^E(x)|T_{ij}\rangle.
\end{equation}
Due to the rotational invariance, the five polarizations give the same
matrix element of the tensor.
\par
The Lorentz invariance is broken on the spacetime lattice. The
zero-momentum stationary glueball state on the lattice must be an
irreducible representation (irrep) of the lattice symmetry group,
i.e. the 24-element octahedral point group $O$. The irreps of
group $O$ are classified as $A_1$, $A_2$, $E$, $T_1$ and $T_2$,
which are the counterparts of angular momentum $J$ for the $SO(3)$
rotational group and have dimensions 1, 1, 2, 3, and 3,
respectively. Together with parity $P$ and charge conjugate $C$
transformations, the full symmetry group of simple cubic lattice
is $O\bigotimes P\bigotimes C$ and the total quantum number of the
lattice glueball state is $R^{PC}$, where $R$ stands for $
A_1,A_2,E,T_1$ or $T_2$, and $PC$ can be $++$,$+-$,$-+$, or $--$.
As mentioned above, we are interested in the matrix elements of
the scalar, pseudoscalar, and tensor operators, which correspond
to the irreps $R^{PC}$ through the following relation:
$0^{++}\leftrightarrow A_1^{++}$, $0^{-+} \leftrightarrow
A_1^{-+}$, and $2^{++} \leftrightarrow E^{++}\oplus T_2^{++}$. The
zero-momentum operators are obtained by summing up all the local
operators on the same time-slice in each channel. The
dimensionless lattice operators are given below explicitly.
\par
The magnetic and electric scalar (labeled $(S,B)$ and $(S,E)$
respectively) belong to the $A_1^{++}$ representation and are defined as
\begin{eqnarray}
\label{def3}
O^{(S,B)}(t) &=& \frac{g^2a_s^4}{L^3}\sum_{{\bf x}_i} {\rm Tr} {\bf
B}^2({\bf x}_i,t),\nonumber\\
O^{(S,E)}(t) &=&\frac{g^2a_s^4}{L^3}\sum_{{\bf x}_i} {\rm Tr} {\bf 
E}^2({\bf
x}_i,t),
\end{eqnarray}
while the pseudoscalar (labeled $(P)$) belongs to the $A_1^{-+}$
representation and is defined as
\begin{equation}
\label{def4}
O^{(PS)}(t) = \frac{g^2a_s^4}{L^3}\sum_{{\bf x}_i}{\rm Tr} {\bf E}({\bf
x}_i,t)\cdot{\bf
B}({\bf x}_i,t).
\end{equation}

In the above questions,  $L^3$ is the dimensionless spatial
volume of the lattice and $a_s$ the spatial lattice spacing. For the
spin-two irrep of
$SO(3)$, the five polarizations are split
across the $E$ and $T_2$ irreps of $O$ and so the tensor
operators must be decomposed into their $E$ and $T_2$ irreducible
contents. The four resulting lattice operators are labeled
$(E,B)$, $(E,E)$, $({T_2},B)$, and $({T_2},E)$, and are
given in terms of their continuum counterparts as
\begin{eqnarray}
\label{def5}
   O_1^{({T_2},B)}(t) &=&\frac{g^2a_s^4}{L^3}\sum_{{\bf x}_i}{\rm Tr}B_2
B_3({\bf
x}_i,t)
\nonumber\\
   O_2^{({T_2},B)}(t) &=&\frac{g^2a_s^4}{L^3}\sum_{{\bf x}_i}{\rm Tr}B_3 
B_1
({\bf
x}_i,t)
\nonumber\\
   O_3^{({T_2},B)}(t) &=&\frac{g^2a_s^4}{L^3}\sum_{{\bf x}_i}{\rm Tr}B_1 
B_2
({\bf
x}_i,t)
\nonumber\\
   O_1^{({T_2},E)}(t) &=&\frac{g^2a_s^4}{L^3}\sum_{{\bf x}_i}{\rm Tr}E_2
E_3({\bf
x}_i,t)
\nonumber\\
   O_2^{({T_2},E)}(t) &=&\frac{g^2a_s^4}{L^3}\sum_{{\bf x}_i}{\rm Tr}E_3 
E_1
({\bf
x}_i,t)
\nonumber\\
   O_3^{({T_2},E)}(t) &=&\frac{g^2a_s^4}{L^3}\sum_{{\bf x}_i}{\rm Tr}E_1
E_2({\bf
x}_i,t)
\end{eqnarray}
\begin{eqnarray}
\label{def6}
   O_1^{(E,B)}(t) &=&\frac{g^2a_s^4}{L^3}\sum_{{\bf 
x}_i}\frac{1}{2}{\rm Tr}
[B_1^2-B_2^2]({\bf x}_i,t)
\nonumber\\
   O_2^{(E,B)}(t) &=&\frac{g^2a_s^4}{L^3}\sum_{{\bf
x}_i}\frac{1}{2\sqrt{3}}{\rm Tr}
[2 B_3^2-B_1^2-B_2^2]({\bf
x}_i,t)\nonumber\\
   O_1^{(E,E)}(t) &=&\frac{g^2a_s^4}{L^3}\sum_{{\bf 
x}_i}\frac{1}{2}{\rm Tr}
[E_1^2-E_2^2]({\bf x}_i,t)
\nonumber\\
   O_2^{(E,E)}(t) &=&\frac{g^2a_s^4}{L^3}\sum_{{\bf
x}_i}\frac{1}{2\sqrt{3}}{\rm Tr}
[2 E_3^2-E_1^2-E_2^2]({\bf x}_i,t)\nonumber\\
\end{eqnarray}
where the coefficients guarantee that the five components are
normalized. Thus we have seven different matrix elements to be
calculated on the lattice,
\begin{eqnarray}
\label{def7}
T(S,B) &=& \langle 0|O^{(S,B)}|A_1^{++}\rangle\nonumber\\
T(S,E) &=& \langle 0|O^{(S,E)}|A_1^{++}\rangle\nonumber\\
T(E,B) &=& \frac{1}{2}\langle
0|O_1^{(E,B)}+O_2^{(E,B)}|E^{++}\rangle\nonumber\\
T(E,E) &=& \frac{1}{2}\langle
0|O_1^{(E,E)}+O_2^{(E,E)}|E^{++}\rangle\nonumber\\
T(T_2,B) &=& \frac{1}{3}\langle
0|O_1^{(T_2,B)}+O_2^{(T_2,B)}+O_3^{(T_2,B)}|T_2^{++}\rangle\nonumber\\
T(T_2,E) &=& \frac{1}{3}\langle
0|O_1^{(T_2,E)}+O_2^{(T_2,E)}+O_3^{(T_2,E)}|T_2^{++}\rangle\nonumber\\
T(PS) &=& \langle 0|O^{(PS)}|A_1^{-+}\rangle
\end{eqnarray}

\par
However, in the practical lattice study, the operators $O^R(t)$ ( 
with $R$ representing the labels mentioned above) are not constructed  
directly by $B_i$ or $E_i$, but by the proper combinations of different 
Wilson loops.  The first definition is based on the linearly 
combination of a set of basic Wilson loops with the requirement that the 
small-$a$ expansion of each combination give the correct continuum form 
shown respectively in Eq.~(\ref{def3})-(\ref{def7}). We call the local 
operators through this definition Type-I operators. The second 
construction is that the lattice version of $B_i$ and $E_i$ are defined 
first by Wilson loops and are used to compose the local operators 
according to Eq.~(\ref{def3})-(\ref{def7}). The local operators through 
this construction is denoted Type-II. 

\par
The following details the constructions of Type-I and Type-II local 
operators on the anisotropic lattice with the spatial lattice spacing 
$a_s$ and the temporal lattice spacing $a_t$. With the tadpole 
improvement~\cite{lepage}, the tree-level Symanzik's improvement scheme 
is implemented to reduce the lattice artifact in defining the local 
operators. Since the aspect ratio of the anisotropic lattice, 
$\xi=a_s/a_t$, is always set to be much larger than 1, the leading
discretization errors in the local operators are at
$O(a_t^2, a_s^4, \alpha_s a_s^2)$ in this work.

\subsection*{A. Type-I Operator}

\begin{figure}
\includegraphics[height=5.0cm]{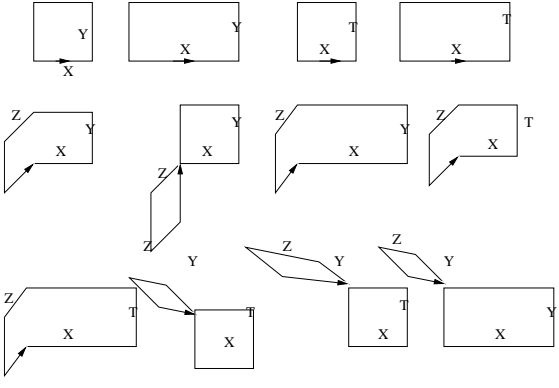}
\caption{\label{prototype} Wilson loops used in making the Type-I
gluonic operators.}
\end{figure}
\begin{table}
\caption{ \label{tab:element} The Wilson loops used in the
construction of the Type-I gluonic operators. They are illustrated
by the ordered paths in the table, where $X$,$Y$,$Z$, and $T$ are
directions and the minus sign indicates the path going in the negative
direction of that axis. ($l_s$,$l_t$) give the numbers of the
spatial and temporal links involved in the Wilson loop. These
numbers also give the powers of the tadpole parameter.
    }
\begin{ruledtabular}
\begin{tabular}{cccc}
Index   & Name      &   Prototype path  & No. of Links\\
 i      &               &    $C_{i,1}$          &($l_s$,$l_t$)\\\hline
 1  & S-Plaquette   &[X,Y,-X,-Y]        &(4,0)        \\
 2  & S-Rectangle   &[X,X,Y,-X,-X,-Y]   &(6,0)        \\
 3  & T-Plaquette   &[X,T,-X,-T]        &(2,2)        \\
 4  & T-Rectangle   &[X,X,T,-X,-X,-T]   &(4,2)        \\
 5  & S-Chair   &[X,Y,-X,Z,-Y,-Z]   &(6,0)        \\
 6  & S-Butterfly   &[X,Y,-X,-Y,Z,-Y,-Z,Y]  &(8,0)        \\
 7  & S-Sunbed  &[X,X,Y,-X,-X,Z,-Y,-Z]  &(8,0)        \\
 8  & T-Chair   &[X,T,-X,Z,-T,-Z]   &(4,2)        \\
 9  & T-Sunbed  &[X,X,T,-X,-X,Z,-T,-Z]  &(6,2)        \\
 10 & Knot      &[X,T,-X,-T,Z,-Y,-Z,Y]  &(6,2)        \\
 11 & LS-Knot   &[X,T,-X,-T,Z,Z,-Y,-Z,-Z,Y] &(8,2)    \\
 12 & LT-Knot   &[X,X,T,-X,-X,-T,Z,-Y,-Z,Y] &(8,2)
\end{tabular}
\end{ruledtabular}
\end{table}

The Type-I operators are constructed from a set of basic Wilson loops 
as illustrated in Fig.~\ref{prototype} and Table~\ref{tab:element}, 
which are chosen by the requirement that the first terms of the small-$a$ 
expansion of these loops give the desired continuum operator forms 
discussed above. We take following steps in the construction. First, 
this set of Wilson loops are acted on by the 24 symmetry operations of 
the cubic point group $O$ (as listed in the Table~\ref{tab:cubic}), 
resulting in 24
copies with different orientations for each type of Wilson loops.
Next, they are linearly combined to realize the irreps of the group $O$.
The coefficients for each irreps, say, $A_1$, $E$, and $T_2$, are listed
in Table~\ref{magic}. To reduce the lattice artifact due to the finite
lattice spacing, the tree level Symanzik's improvement is used, which
means that differently shaped Wilson loops are combined to construct one
operator so that the lattice artifacts are pushed to higher order of
lattice spacing. Tadpole improvement is also implemented to
improve the reliability of the lattice spacing expansion at the tree
level~\cite{lepage}.
\begin{table}
\caption{  \label{tab:cubic}
The 24 symmetric operations of the cubic point group $O$ are represented
by the coordinate transformation.
        }
\begin{ruledtabular}
\begin{tabular}{cl|cl}
Index& ~~~~~~~Operation &Index  &~~~~~~~Operation\\
\hline
1 &$(x,y,z)\rightarrow (x,y,z)$&
13&$(x,y,z)\rightarrow (-z,y,x)$\\
2 &$(x,y,z)\rightarrow (-z,-y,-x)$&
14&$(x,y,z)\rightarrow (-x,-y,z)$\\
3 &$(x,y,z)\rightarrow (z,x,y)$&
15&$(x,y,z)\rightarrow (-y,x,z)$\\
4 &$(x,y,z)\rightarrow (-y,-x,-z)$&
16&$(x,y,z)\rightarrow (-z,-x,y)$\\
5 &$(x,y,z)\rightarrow ( y,z,x)$&
17&$(x,y,z)\rightarrow (-x,z,y)$\\
6 &$(x,y,z)\rightarrow (-x,-z,-y)$&
18&$(x,y,z)\rightarrow (-y,-z,x)$\\
7 &$(x,y,z)\rightarrow ( z,y,-x)$&
19&$(x,y,z)\rightarrow (-x,y,-z)$\\
8 &$(x,y,z)\rightarrow (x,-y,-z)$&
20&$(x,y,z)\rightarrow (z,-y,x)$\\
9 &$(x,y,z)\rightarrow (y,x,-z)$&
21&$(x,y,z)\rightarrow (-z,x,-y)$\\
10&$(x,y,z)\rightarrow (z,-x,-y)$&
22&$(x,y,z)\rightarrow (y,-x,z)$\\
11&$(x,y,z)\rightarrow (x,z,-y)$&
23&$(x,y,z)\rightarrow (-y,z,-x)$\\
12&$(x,y,z)\rightarrow (y,-z,-x)$&
24&$(x,y,z)\rightarrow (x,-z,y)$
\end{tabular}
\end{ruledtabular}
\end{table}
\begin{table*}
\caption{\label{magic}Combinational coefficients used in
Eq.~(\ref{realpart}) to construct the irreps of the cubic point
group. }
\begin{ruledtabular}
\begin{tabular}{cccccccccc}
$r\in O$&
$\alpha_r^{A_1}$&
$\alpha_r^{E^+  }(1,1)$&
$\alpha_r^{E^+}(1,2)$&
$\alpha_r^{E^+}(2,1)$&
$\alpha_r^{E^+}(2,2)$&
$\alpha_r^{T_2^+}(2,1)$&
$\alpha_r^{T_2^+}(2,2)$&
$\alpha_r^{T_2^+}(2,3)$\\
\hline
 1&1&-1&$-\frac{1}{\sqrt{3}}$&$-\frac{1}{\sqrt{3}}$& 1& 0& 1& 0\\
 2&1&-1&$-\frac{1}{\sqrt{3}}$&$ \frac{1}{\sqrt{3}}$&-1& 0& 1& 0\\
 3&1& 1&$-\frac{1}{\sqrt{3}}$&$-\frac{1}{\sqrt{3}}$&-1& 0& 0& 1\\
 4&1& 1&$-\frac{1}{\sqrt{3}}$&$ \frac{1}{\sqrt{3}}$& 1& 0& 0& 1\\
 5&1& 0&$ \frac{2}{\sqrt{3}}$&$ \frac{2}{\sqrt{3}}$& 0& 1& 0& 0\\
 6&1& 0&$ \frac{2}{\sqrt{3}}$&$ \frac{2}{\sqrt{3}}$& 0& 1& 0& 0\\
 7&1&-1&$-\frac{1}{\sqrt{3}}$&$ \frac{1}{\sqrt{3}}$&-1& 0&-1& 0\\
 8&1&-1&$-\frac{1}{\sqrt{3}}$&$-\frac{1}{\sqrt{3}}$& 1& 0&-1& 0\\
 9&1& 1&$-\frac{1}{\sqrt{3}}$&$ \frac{1}{\sqrt{3}}$& 1& 0& 0&-1\\
10&1& 1&$-\frac{1}{\sqrt{3}}$&$-\frac{1}{\sqrt{3}}$&-1& 0& 0&-1\\
11&1& 0&$ \frac{2}{\sqrt{3}}$&$-\frac{2}{\sqrt{3}}$& 0&-1& 0& 0\\
12&1& 0&$ \frac{2}{\sqrt{3}}$&$ \frac{2}{\sqrt{3}}$& 0&-1& 0& 0\\
13&1&-1&$-\frac{1}{\sqrt{3}}$&$ \frac{1}{\sqrt{3}}$&-1& 0&-1& 0\\
14&1&-1&$-\frac{1}{\sqrt{3}}$&$-\frac{1}{\sqrt{3}}$& 1& 0&-1& 0\\
15&1& 1&$-\frac{1}{\sqrt{3}}$&$ \frac{1}{\sqrt{3}}$& 1& 0& 0&-1\\
16&1& 1&$-\frac{1}{\sqrt{3}}$&$-\frac{1}{\sqrt{3}}$&-1& 0& 0&-1\\
17&1& 0&$ \frac{2}{\sqrt{3}}$&$-\frac{2}{\sqrt{3}}$& 0&-1& 0& 0\\
18&1& 0&$ \frac{2}{\sqrt{3}}$&$ \frac{2}{\sqrt{3}}$& 0&-1& 0& 0\\
19&1&-1&$-\frac{1}{\sqrt{3}}$&$-\frac{1}{\sqrt{3}}$& 1& 0& 1& 0\\
20&1&-1&$-\frac{1}{\sqrt{3}}$&$ \frac{1}{\sqrt{3}}$&-1& 0& 1& 0\\
21&1& 1&$-\frac{1}{\sqrt{3}}$&$-\frac{1}{\sqrt{3}}$&-1& 0& 0& 1\\
22&1& 1&$-\frac{1}{\sqrt{3}}$&$ \frac{1}{\sqrt{3}}$& 1& 0& 0& 1\\
23&1& 0&$ \frac{2}{\sqrt{3}}$&$ \frac{2}{\sqrt{3}}$& 0& 1& 0& 0\\
24&1& 0&$ \frac{2}{\sqrt{3}}$&$-\frac{2}{\sqrt{3}}$& 0& 1& 0& 0
\end{tabular}
\end{ruledtabular}
\end{table*}

\par
Apart from the rotational symmetry, the constructed operators
should have also definite parity and charge conjugation
properties. The symmetric/antisymmetric combination of a Wilson loop
and its parity-transformed counterparts gives the positive/negative
parity. The $C=+$ operators can be realized by taking the real part
of a Wilson loop.
 \par
Any operator that includes the chromo-electric field will involve
loops with finite extent in the time direction. Operators must be
defined for a chosen value of $t$. The simplest way to enforce
this definition is to ensure the operators on time-slice $t$ are
eigenstates of the time reversal operators, $T$ about that time
slice. The eigenvalue of this reversal must be the same as the
parity of the field operator to ensure the correct dimension-four
operator is reproduced. The chromo-electric scalar and tensor
operators transform positively under $T$, while the pseudoscalar
transforms negatively. The combination coefficients of the time-reversed
loops are the products of coefficients of original loops and the
time reversal eigenvalues of the time-reversed operator.
\par
It should be noted that the combination coefficients are
independent of the shape of loops and each irreducible
representation corresponds to a specific set of combination
coefficients, denoted by $\alpha_r^{(R)}, r = 1,2,\ldots,24$.
\par
For clarity, we give the explicit formula of the thirteen zero-momentum
gluonic operators as follows:
\begin{eqnarray}
\label{realpart} O^{(S,B)}(t) &=&
\sum\limits_{r,x}(\frac{5}{12 
u_s^4}{\rm ReTr}[1-U_{C_{1,r}}(x,t)]\nonumber\\
&-&\frac{1}{24 u_s^6}{\rm ReTr}[1-U_{C_{2,r}}(x,t)])\nonumber\\
O^{(S,E)}(t) &=& {\xi^2} \sum\limits_{r,x}(\frac{1}{12 u_s^2
u_t^2}{\rm ReTr}[1-U_{C_{3,r}}(x,t)]\nonumber\\
&-&\frac{1}{192 u_s^4 u_t^2}{\rm ReTr}[1-U_{C_{4,r}}(x,t)])\nonumber\\
O_i^{(T_E,B)}(t) &=& \sum\limits_{r,x}
\{\alpha_r^{E^+}(1,i)(\frac{5}{12 
u_s^4}{\rm ReTr}U_{C_{1,r}}(x,t)\nonumber\\
&-&\frac{1}{96 u_s^6}{\rm ReTr}U_{C_{2,r}}(x,t) )\nonumber\\
&+&\alpha_r^{E^+}(2,i)\frac{\sqrt{3}}{96
u_s^6}{\rm ReTr}U_{C_{2,r}}(x,t)\}\nonumber \\
O_i^{(T_E,E)}(t) &=&\xi^2
\sum\limits_{r,x}\alpha_r^{E^+}(1,i)(\frac{1}{6 u_s^2
u_t^2}{\rm ReTr}U_{C_{3,r}}(x,t)\nonumber \\
&-&\frac{1}{96u_s^4 u_t^2}{\rm ReTr}U_{C_{4,r}}(x,t))\nonumber\\
O_i^{(T_{T_2},B)}(t) &=&  \sum\limits_{r,x}\alpha_r^{T_2^+}(2,i)(
  -\frac{11}{96 u_s^6} {\rm ReTr}U_{C_{5,r}}(x,t)\nonumber\\
&+&\frac{1}{96 u_s^8}  {\rm ReTr}U_{C_{6,r}}(x,t)
  +\frac{1}{48 u_s^8}  {\rm ReTr}U_{C_{7,r}}(x,t) )\nonumber\\
O_i^{(T_{T_2},E)}(t) &=& {\xi^2}
\sum\limits_{r,x}\alpha_r^{T_2^+}(2,i)(
 \frac{5}{96 u_s^4 u_t^2}{\rm ReTr}U_{C_{8,r}}(x,t)\nonumber\\
&-&\frac{1}{96 u_s^6 u_t^2}{\rm ReTr}U_{C_{9,r}}(x,t) )\nonumber\\
O^{(PS)}(t) &=& \xi\sum\limits_{r,x}\alpha_r^{A_1^-} ( \frac{6}{96
u_s^6
u_t^2}{\rm ReTr}U_{C_{10,r}}(x,t)\nonumber\\
 &-&\frac{1}{96 u_s^8
u_t^2}{\rm ReTr}U_{C_{11,r}}(x,t)\nonumber\\
 &-&\frac{1}{192 u_s^8
u_t^2}{\rm ReTr}U_{C_{12,r}}(x,t) ),
\end{eqnarray}
where $U_{C_{i,r}}(x,t)$ is the loop generated by operating $r$-th
rotation on the $i$-th prototype loop in Table~\ref{tab:element}.
The tadpole parameters $u_s$ and $u_t$ here come from the
renormalization of spatial and
temporal gauge links, respectively: $U_j(x)\rightarrow U_j(x)/u_s$ and
$U_t(x)\rightarrow U_t(x)/u_t$. The determination of $u_s$ and $u_t$ is
described in Section III.

\subsection*{B. Type-II Operators}

Generally speaking, there can be many ways to define the lattice
local operators as long as these definitions give the same continuum
limit, and some may have smaller lattice artifacts at finite lattice 
spacing
compared to others. This
motivates us to design another type of lattice gluonic operators,
called Type-II operators in this work, apart from the construction
described above. Both types of operators are used for
a self-consistent check.
\par
\begin{figure}[htb!]
\includegraphics[height=5.0cm]{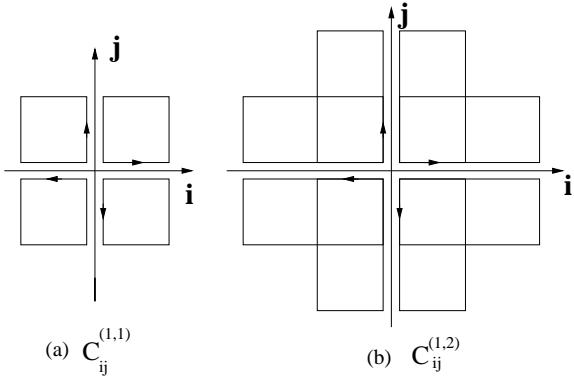}
\caption[]{\label{clover} The clover-shape combinations of spatial
plaquette and rectangle, which are used to derive the gauge field
strength $F_{\mu\nu}$ on the lattice. Here $i$ and $j$ are the
indices of the spatial direction. }
\end{figure}

According to the non-Abelian Stokes theorem~\cite{stokes}, a rectangle
Wilson loop
$P_{\mu\nu}^{a\times b}(x)$ of size $a\times b$, with $a,b$ small,
can be expanded as
\begin{widetext}
\begin{eqnarray}
\label{expansion}
P_{\mu\nu}^{a\times b}(x) &=& {\bf 1} +
ab\left(F_{\mu\nu}(x)+\frac{1}{2}(aD_{\mu}+bD_\nu)F_{\mu\nu}(x)
+\frac{1}{12}(2a^2D_\mu ^2
               +3abD_\mu D_\nu
               +2b^2D_{\nu}^2)F_{\mu\nu}(x)\right. \nonumber\\
&+&
\left.
\frac{1}{24}(a^3D_{\mu}^3 + 
                2a^2 b D_{\mu}^2 D_{\nu} + 
                2 ab^2D_{\mu}D_{\nu}^2 + b^3 D_{\nu}^3)F_{\mu\nu}(x)
\right)\nonumber\\
&+&(ab)^2\left(
\frac{1}{2}F_{\mu\nu}^2(x) +
\frac{1}{2}F_{\mu\nu}(x)(aD_{\mu}+bD_{\nu})F_{\mu\nu}(x)
+\frac{1}{24}F_{\mu\nu}(x)(a^2 D_{\mu}^2+b^2
D_{\nu}^2)F_{\mu\nu}(x)\right)\nonumber\\
&+&\frac{1}{6}(ab)^3 F_{\mu\nu}^3 + O((ab)^4).
\end{eqnarray}
\end{widetext}
For simplicity, the factor $ig$ is absorbed into the quantity
$F_{\mu\nu}$ and will be reconsidered when comparing with the continuum
form.  This expression can be simplified by the clover-type combination
which is defined as
\begin{eqnarray}
C_{\mu\nu}^{(a,b)}&(x)& = \frac{1}{8}\left[ \right.\nonumber\\
&&\left(P_{\mu\nu}^{a\times b}(x)+P_{\nu-\mu}^{a\times b}(x)
      +P_{-\mu-\nu}^{a\times b}(x)+P_{-\nu\mu}^{a\times b}(x)
\right)
\nonumber\\
&+&\left(P_{\mu\nu}^{b\times
a}(x)+P_{\nu-\mu}^{b\times a}(x)+P_{-\mu-\nu}^{b\times
a}(x)+P_{-\nu\mu}^{b\times a}(x)\right) 
\nonumber\\
&&\left.\right]
\end{eqnarray}
where
\begin{equation}
P_{\pm\mu\pm\nu}(x) = U_{\pm\mu}(x)U_{\pm\nu}(x\pm a
\mu)U_{\pm\mu}^{\dagger}(x\pm b \nu) U_{\pm\nu}^{\dagger}(x).
\end{equation}
The small $ab$ expansion of $ P_{\pm\mu\pm\nu}(x)$ is similar to
Eq.~(\ref{expansion}) by replacing $a$ and $b$ with $\pm a$ and
$\pm b$, respectively. This clover-type combination is illustrated
in Fig.~\ref{clover}.

As a result, the tree-level expansion of
$C_{\mu\nu}(x)$ is explicitly derived as
\begin{eqnarray}
{\rm Re}\,C_{\mu\nu}(x) &=& {\bf 1} + \frac{1}{2}(ab)^2
F_{\mu\nu}^2(x)\nonumber\\
&+&\frac{1}{48}(ab)^2 (a^2+b^2)F_{\mu\nu}(x)(D_{\mu}^2+
D_{\nu}^2)F_{\mu\nu} \nonumber\\
&+& O((ab)^4),\nonumber\\
{\rm Im}\,C_{\mu\nu}(x) &=& abF_{\mu\nu}(x) \nonumber\\
&+& \frac{1}{12}ab(a^2+b^2)(D_{\mu}^2+D_{\nu}^2)F_{\mu\nu}(x)
\nonumber\\
&+&\frac{1}{6}(ab)^3 (F_{\mu\nu})^3 + \ldots.
\end{eqnarray}

Using ${\rm Im}\,C_{\mu\nu}(x)$ with different $a$ and $b$ as the 
elementary
components, the local operators $F_{\mu\nu}F_{\rho\sigma}$ can be
defined through the lattice version of the gauge field
strength tensor
\begin{eqnarray}
\hat{F}_{\mu\nu}(x) &=& \frac{1}{3}\left(5 {\rm Im}\,C_{\mu\nu}^{1,1}(x)
            - {\rm Im}\,C_{\mu\nu}^{1,2}(x)\right)\nonumber\\
&=& a^2\left( F_{\mu\nu}(x)
-\frac{1}{6}a^4(F_{\mu\nu}(x))^3+O(a^6)\right),\nonumber \\
\end{eqnarray}
which is improved up to $O(a^4)$ by the combination of the
clover-leaf diagram in Fig.~\ref{clover}(a) and the wind-mill
diagram in Fig.~\ref{clover}(b). Thus the fourth-dimensional gauge
operator $F_{\mu\nu}F_{\rho\sigma}$ can be derived from
$\hat{F}_{\mu\nu}(x)\hat{F}_{\rho\sigma}(x)$,
 \begin{eqnarray}
\label{imag1}
\hat{F}_{\mu\nu}(x)\hat{F}_{\rho\sigma}(x)&=&a^4
F_{\mu\nu}(x)F_{\rho\sigma}(x) \nonumber\\
&-&\frac{1}{6}a^8 F_{\mu\nu}
    \left(F_{\mu\nu}^2(x)+F_{\rho\sigma}^2(x)\right)F_{\rho\sigma}(x)
\nonumber\\
    &+&\ldots.
\end{eqnarray}
\par
In order to check the effectiveness of the improvement scheme of
lattice local operators, we have also derived another lattice
definition of $F_{\mu\nu}F_{\rho\sigma}(x)$, which is through the
direct combination of ${\rm Im}\,C_{\mu\nu}(x)$, say,
\begin{widetext}
\begin{eqnarray}
\label{imag2}
\widehat{F_{\mu\nu}(x)F_{\rho\sigma}(x)}&=&\frac{1}{3}
\left(7 {\rm Im}\,C_{\mu\nu}^{1,1}(x)\cdot
{\rm Im}\,C_{\rho\sigma}^{1,1}(x)
         - {\rm Im}\,C_{\mu\nu}^{1,1}(x)\cdot 
{\rm Im}\,C_{\rho\sigma}^{1,2}(x)
         - {\rm Im}\,C_{\mu\nu}^{1,2}(x)\cdot 
{\rm Im}\,C_{\rho\sigma}^{1,1}(x)
       \right) \nonumber\\
& =& a^4 F_{\mu}{\nu}(x)F_{\rho\sigma}(x) -
        \frac{1}{6}a^8 F_{\mu\nu}
    \left(F_{\mu\nu}^2(x)+F_{\rho\sigma}^2(x)\right)F_{\rho\sigma}(x)
        -\frac{1}{36}a^8 (D_{\mu}^2+D_{\nu}^2)F_{\mu\nu}
                         (D_{\rho}^2+D_{\sigma}^2)F_{\rho\sigma} + \ldots
\nonumber \\
\end{eqnarray}
\end{widetext}
One can find that, at tree level, the lowest order difference
between these two definitions, denoted by $\Delta$, is
\begin{equation}
\Delta = -\frac{1}{36}a^8 (D_{\mu}^2+D_{\nu}^2)F_{\mu\nu}
                         (D_{\rho}^2+D_{\sigma}^2)F_{\rho\sigma} + \ldots.
\end{equation}

In the practical calculation, based on these two approaches, we
construct two version of operators in $T_2^{++}$ irreps.
In Section IV, we find the different definitions result in about
3--4\% discrepancy for the measured matrix elements.
\par
The discussion above are based on the classical (or tree level)
series expansion of Wilson loops. For this to be reliable, the
tadpole improvement should be applied, which means the tadpole
parameter should be included in the above expressions.
Specifically, a $n$-link spatial Wilson loop $C_{\mu\nu}(x)$ should be
divided by a tadpole factor $1/u_s^n$. From the data
analysis to be shown in Sec.~\ref{details} that the tadpole improvement
alleviates most of the dependence on finite lattice spacing.
The improvement scheme of the local operators corresponding to $E_iE_j$
is a little different from that of $B_iB_j$.
 Since $a_t\ll a_s$, all the temporal Wilson loops
included in the improvement have only one lattice spacing
extension in the time direction. In other words, for the temporal
loops, we do not include the windmill diagrams which involve
 two lattice spacing in the time direction. Thus the combination
coefficients in the above expression are modified accordingly. We omit
the explicit expression here.

\section*{III. Numerical Details} \label{details}

Since the implementation of the tadpole improved gauge action on
anisotropic lattices was verified to be very successful and
efficient in the determination of the glueball spectrum~\cite{old3},
we use the same techniques to calculate the glueball matrix
elements. We adopt the anisotropic gauge action used by
Morningstar and Peardon in{}~\cite{old3}
\begin{equation}  \label{action}
S = \beta\large\{\frac{5}{3}\frac{\Omega_{sp}}{\xi u_s^4} +
                 \frac{4}{3}\frac{\xi\Omega_{tp}}{u_s^2 u_t^2} -
                 \frac{1}{12}\frac{\Omega_{sr}}{\xi u_s^6} -
                 \frac{1}{12}\frac{\xi\Omega_{str}}{u_s^4 u_t^2}\large\},
\end{equation}
where $\beta = 6/g^2$, $g$ is the QCD coupling constant, $\xi$ is
the aspect ratio for anisotropy, $u_s$ and $u_t$ are the tadpole
improvement parameters of spatial and temporal gauge links,
respectively, and $\Omega_C$'s are the sums of various Wilson
loops over the total lattice (the explicit expression can be found
in Ref.~\cite{old3}.) In practice, $u_s$ is defined by the
expectation value of the spatial plaquette, namely, $u_s =
(\langle 1/3 Tr P_{ss'}\rangle)^{1/4}$, and $u_t$ is set to 1.
Theoretically, the bare anisotropy $\xi$ should be finely tuned to
give the correct physical anisotropy $\xi_{phys} = a_s/a_t$, but
in our practical case $\xi$ is always taken as the same as
$\xi_{phys}$ because the discrepancy is shown to be within $1-2$
percents when the improved action in Eq.~(\ref{action}) is
used~\cite{old3}. For each coupling constant $\beta$ and $\xi$,
$u_s$ is determined self-consistently in the Monte Carlo updating.

\par
\begin{figure}
\includegraphics[height=5.0cm]{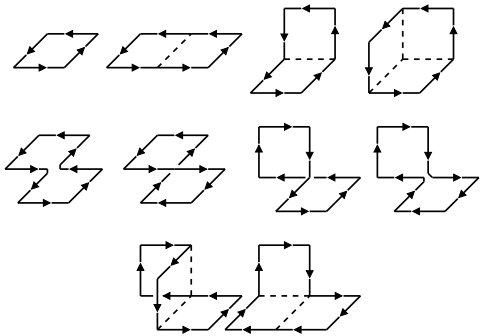}
\caption{\label{proto} Wilson loops used in making the smeared
glueball operators.}
\end{figure}

The gauge configurations were generated by using Cabibbo-Marinari(CM)
pseudo-heatbath and the SU(2) subgroup micro-canonical
over-relaxation (OR) methods. Three compound sweeps were performed
between measurements, where a compound sweep is made up of one CM
updating sweep followed by 5 OR sweeps. The measurements of
$n_{mb}$ configurations are averaged in each bin, and $n_{bin}$ bins are obtained.
Table~\ref{tab:lattice} lists the relevant input parameters
for lattices with 5 different $\beta$. For the case of
$\beta = 2.4$, there are three lattice volumes to study the finite
volume effects.
\begin{table}
\caption{\label{tab:lattice} The input parameters for the calculation.
Values for the coupling $\beta$, anisotropy $\xi$, the tadpole
parameter $u_s^4$, the single-link smearing parameter $\lambda_s$,
the double-link smearing parameter $\lambda_f$, lattice size, and
the number of measurements are listed.}
\begin{ruledtabular}
\begin{tabular}{cccccrr}
$\beta$ &  $\xi$  &  $u_s^4$  & $\lambda_s$ & $\lambda_f$ &
 $L^3\times T$ & $N_{conf} (n_{bins}\times n_{mb}$) \\\hline
   2.4  & 5 & 0.409 & 0.1 & 0.5 & $8^3\times 40$ & $999\times 100$ \\
        &   &       &     &     &$12^3\times 64$ & $100\times 100$ \\
        &   &       &     &     &$16^3\times 80$ & $100\times 100$ \\
   2.6  & 5 & 0.438 & 0.1 & 0.5 &$12^3\times 64$ & $86 \times 100$ \\
   2.7  & 5 & 0.451 & 0.1 & 0.5 &$12^3\times 64$ & $100\times 100$ \\
   3.0  & 3 & 0.500 & 0.4 & 0.5 &$16^3\times 48$ & $100\times 100$ \\
   3.2  & 3 & 0.521 & 0.4 & 0.5 &$24^3\times 72$ & $ 79\times 100$ \\
\end{tabular}
\end{ruledtabular}
\end{table}
In order to calculate the matrix element such as $\langle
0|O|G\rangle$, it is desirable to have the glueball state $ |G\rangle$
determined as precisely as possible. In this work, the glueball
states $|G\rangle$ with quantum number $R = A_1^{PC}$, $A_2^{PC}$,
$E^{PC}$, $T_1^{PC}$, and $T_2^{PC}$ are generated by smeared
gluonic operators $O_S^{R}$, which are constructed by exploiting
link-smearing and variational techniques in a sequence of three
steps outlined below. First, for each generated gauge
configuration, we perform six smearing/fuzzing schemes to the spatial
links, which are various combinations of the single-link procedure
\begin{widetext}
\begin{eqnarray}  \label{s/f}
U_j^s(x) &=& P_{SU(3)}\{U_j(x)+\lambda_s\sum\limits_{\pm(k\neq j)}
U_k(x)U_j(x+\hat{k})U_k^{\dagger}(x+\hat{j})\},\nonumber \\
U_j^f(x) &=&
P_{SU(3)}\{U_j(x)U_j(x+\hat{j})+\lambda_f\sum\limits_{\pm(k\neq
j)}U_k(x)U_j(x+\hat{k})U_j(x+\hat{j}+\hat{k})U_k(x+2\hat{j})\},
\end{eqnarray}
\end{widetext}
where $P_{SU(3)}$ denotes the projection into $SU(3)$ and is
realized by the Jacobi method~\cite{liu2} in this work. The six schemes
are
given explicitly as $ s_{\lambda_s}^2$, $ s_{\lambda_s}^4$, $
s_{\lambda_s}^6$, $ f_{\lambda_f}\bigotimes s_{\lambda_s}^2$, $
f_{\lambda_f}\bigotimes s_{\lambda_s}^4$, $ f_{\lambda_f}\bigotimes
s_{\lambda_s}^6$, where $s/f$ denotes smearing/fuzzing procedure defined
in Eq.~(\ref{s/f}). $\lambda_s$ and $\lambda_f$ are tunable
parameters for smearing and fuzzing and take the optimal value
$\lambda_s = 0.1$ ( $0.4$ at $\xi = 3$) and $ \lambda_f = 0.5$. Secondly,
for each smearing/fuzzing scheme, we use ten Wilson loops illustrated in
Fig.~\ref{proto}, which are the same as used in{}~\cite{old3},
as prototypes to construct the operator $\phi_{\alpha}^{(R)}(t)$
which is a linear combination of different oriented spatial loops,
invariant under spatial transformation, and transforms according
to the irreps $R$. There are four independent constructions for
each $R$ (except for $A_2^{-+}$ ) for each smearing/fuzzing scheme, thus 
the
operator $O_S^{(R)}(t)$ is a linear combination of 24
operators, $O_S^{(R)} =\sum\limits_{\alpha} v_\alpha^{(R)}
\phi_{\alpha}^{(R)}(t)$. The coefficients
$v_\alpha^{(R)}$ are determined by a variational method so that
$O_S^{(R)}$ projects mostly to a specific glueball states $|G\rangle$.

What we obtain from the MC simulation are the $24\times 24$ correlation
matrix
\begin{equation}
\tilde{C}_{\alpha\beta}(t) = \sum\limits_{\tau}\langle
0|\bar{\phi}_\alpha^{(R)}(t+\tau)\bar{\phi}_\beta^{(R)}(\tau)|0\rangle,
\end{equation}
where the vacuum subtraction
\begin{equation}
\bar{\phi}_\alpha^{(R)}(t)=\phi_\alpha^{(R)}(t)-\langle
0|\phi_\alpha^{(R)}(t)|0\rangle 
\end{equation}
is only applied to the $A_1^{++}$ channel
which has a vacuum expectation value.
The coefficients are determined in the data analysis stage by
minimizing the effective mass
\begin{equation}
\label{mass}
\tilde{m}(t_D)=-\frac{1}{t_D}\ln\frac
{\sum\limits_{\alpha\beta}
v_\alpha^{(R)}v_\beta^{(R)}\tilde{C}_{\alpha\beta}(t_D)}
{\sum\limits_{\alpha\beta}
v_\alpha^{(R)}v_\beta^{(R)}\tilde{C}_{\alpha\beta}(0)},
\end{equation}
where the time separation for optimization is fixed to $t_D=1$. This is
equivalent to solving the generalized eigenvalue equation
\begin{equation}
\label{eigen}
\tilde{C}(t_D){\bf v}^{(R)} = e^{-t_D\tilde{m}(t_D)}\tilde{C}(0){\bf
v}^{(R)}.
\end{equation}
and the eigenvector ${\bf v}_0^{(R)}$, corresponding to the lowest
effective mass $\tilde{m}_0(t_D)$, yields the coefficients
$v_{0\alpha}^{(R)}$ for the operator $O_S^{(R)}(t)$ which, under
ordinary circumstances, best overlaps with the lowest-lying glueball
$G_0$ in the channel of interest. The operators most overlapping
with the excited states can also be constructed accordingly.
\par
After the smeared operators are determined, the glueball masses and the
matrix elements we are concerned with can be extracted by fitting two
two-point functions, namely, the smeared-smeared correlation function,
\begin{eqnarray}
\label{GG}
C_{SS}(t) &=& \sum\limits_{\tau}\langle
0|O_S(t+\tau)O_S(\tau)|0\rangle\nonumber\\
&=&\sum\limits_n\frac{1}{2M_nV}
\langle 0|O_S|n\rangle \langle n|O_S|0\rangle
e^{-m_n t}\nonumber\\
&\sim& \frac{1}{2m_G V}
|\langle G|O_S|0\rangle|^2 e^{-M_G t}~~(t\rightarrow
\infty),
\end{eqnarray}
and the smeared-local one
\begin{eqnarray}
\label{GM}
C_{SL}(t) &=& \sum\limits_{\tau}\langle
0|O_S(t+\tau)O_L(\tau)|0\rangle\nonumber\\
&=&\sum\limits_n \frac{1}{2M_nV}
\langle 0|O_S|n\rangle \langle n|O_L|0\rangle
e^{-m_n t}\nonumber\\
&\sim& \frac{1}{2M_G V}
\langle G|O_S|0\rangle\langle 0|O_L|G\rangle e^{-M_G t}~~(t\rightarrow
\infty).
\nonumber\\
\end{eqnarray}
$C_{SS}(t)$ and $C_{SL}(t)$ can be fitted simultaneously with three
parameters, i.e., $\langle 0|O_S|G\rangle$, $\langle 0|O_L|G\rangle$, and
the ground state glueball mass $M_G$. $\langle 0|O_L|G\rangle$ is the
glueball matrix element we would like to obtain. Of course, before the physical
results can be derived, a proper renormalization scheme should be performed.
\par
In the following, we describe the calculation details step by step.

\subsection*{A. Setting the Scale Using the Static Potential}
\begin{table}
\caption{\label{tab:spacing} The lattice spacings determined by
$r_0^{-1}=410(20)$ MeV at different $\beta$. }
\begin{ruledtabular}
\begin{tabular}{cccc}
$\beta$ &  $r_0/a_s$  & $a_s/r_0$ & $a_s$ (fm)\\
\hline
2.4 &  2.17(1)    & 0.461(2)  & 0.222(1)  \\
2.6 &  2.74(2)    & 0.365(2)  & 0.176(1)  \\
2.7 &  3.09(2)    & 0.326(2)  & 0.156(1)  \\
3.0 &  4.05(4)    & 0.247(2)  & 0.119(1)  \\
3.2 &  4.76(5)    & 0.210(2)  & 0.101(1)
\end{tabular}
\end{ruledtabular}
\end{table}

The five lattice spacings $a_s$ are determined by calculating the heavy
quark static potential. This part of the calculation is independent of the
production runs. The static-quark potential $V({\bf r})$ can be
extracted from the averages of Wilson loops $W({\bf r},t)$,
\begin{equation}
 W({\bf r},t) = Z({\bf r}) \exp(-tV({\bf r}) ) + \ldots
\end{equation}
which can be measured precisely on the lattice. For each $\beta$,
200 configurations are generated, each of which is
separated by 30 compound sweeps so that the auto-correlation
effects are reduced. Secondly, these configuration are fixed to
temporal gauge so that the Wilson loops can be calculated more
easily. Different smearing schemes are applied to Wilson loops
with different size to increase the overlapping with the ground
states. Finally, the static potential is fitted by the model
\begin{equation}
V({\bf r}) = V_0 + \sigma r +\frac{e_c}{r},
\end{equation}
with three parameters $V_0$, $\sigma$, and $e_c$ in the correlated
fit method. The lattice spacing $a_s$ is determined by
\begin{equation}
\frac{a_s}{r_0}=\sqrt{\frac{\sigma a_s^2}{1.65+e_c}},
\end{equation}
according to the relation $ r^2dV({\bf
r})/dr]_{r=r_0}=1.65$, where $r_0$ is hadronic scale parameter~\cite{sommer94}. The
lattice spacings for different $\beta$ are listed in
Table~\ref{tab:spacing}.

\subsection*{B. The Glueball Mass Spectrum}  \label{glueball_mass}

Glueball masses can be obtained by fitting the two-point
functions $C_{SS}(t)$ directly. It is taken as a testimony of the
effectiveness of the improvement and smearing scheme~\cite{old3}. Even though
we are interested in the glueball matrix elements of the four channels
$A_1^{++}$, $A_1^{-+}$, $E^{++}$, and $T_2^{++}$ in this work,
glueball masses of all the 20 channels $R^{PC}$ are calculated as a by-product.
\par
After the implementation of variational-optimization in each channel, we
can obtain a specific operator which overlaps most with the a specific
state and has little contaminations from other states with the same
$J^{PC}$, so that we can use a single mass term
\begin{equation}
C_{SS}(t) = Z \left( e^{-Mt} + e^{-M(T-t)}\right),
\end{equation}
to fit the two-point function in a time range $t_{min},
\ldots, t_{max}$, which can be determined by observing the
effective mass plateau. As a convention in this work, we use $M_G$ to 
represent the mass of a glueball state in the physical units and 
$M$ to represent the dimensionless mass parameter in the data processing
with the relation $M=M_G a_t$. Generally speaking, for most channels in
each of the five $\beta$ cases, the overlap of the specific
operators on the ground states are all larger than 90\%.
\par
Before performing the continuum extrapolation, we check the finite
volume effects of glueball masses.
Three independent calculations at $\beta=2.4$, $\xi = 5$ were carried out
on a $8^3\times 40$ lattice, a $12^3\times 64$, and a $16^3\times 80$
lattice. These lattices
have spatial volumes of $(1.76~{\rm fm})^3$, $(2.64~{\rm fm})^3$, and
$(3.52~{\rm fm})^3$, respectively, with the lattice spacing
$a_s\sim 0.22\,{\rm fm}$ from $r_0^{-1} = 410(20)$ MeV. For these three 
runs,
all the input parameters are the same except the different lattice
volume.
\par
\begin{figure}[htb!]
\includegraphics[height=7.5cm]{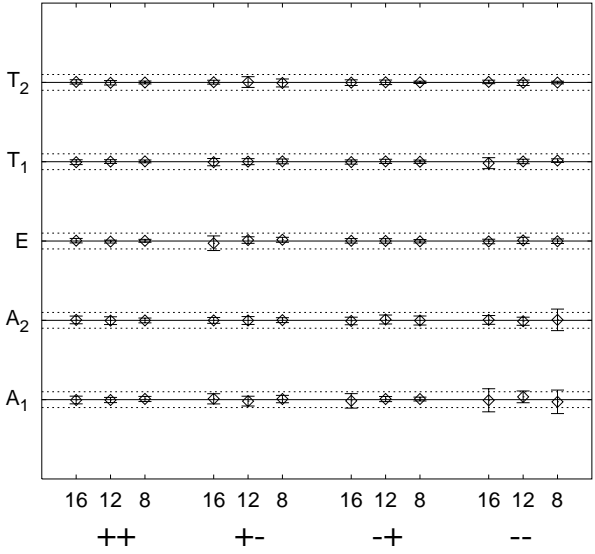}
\caption[]{\label{fvm} Finite-volume effects of glueball masses at
$\beta=2.4$,
$\xi=5$. Each point shows the fractional change $\delta_G(L) =
1-M(L)/\bar{M}$ of the glueball mass, where
$M(L)$ is the glueball mass measured on the lattice $L^3\times
T$ with $L = 8,12,$ and 16, and $\bar{M}$ is the average values over
those from different lattices. The errorbars come from the statistical
errors of $M(L)$. The lattice labels $L$ are shown along the
horizontal axis, and the labels along the vertical axis are the irreps
of lattice symmetry group. The solid lines indicates
$\delta_G=0$, and the dotted lines above the solid lines
indicates $\delta_G = 0.02$, and the dotted line lines below the solid
lines indicate $\delta_G = -0.02$.}
\end{figure}

The calculated glueball mass spectra are listed in
Table~\ref{fve_mass}, where one can find at a glance that the
finite volume effects are very small and in most case the changes
are within errors and consistent with zero statistically. More
precisely, we use the following scheme to illustrate the finite volume 
effect (FVE) quantitatively. Let $\bar{M}$ denote the average value of 
the
glueball masses from the three lattice volumes, $M(L)$ denotes
the glueball mass measured on lattice $L^3\times T$. The
fractional change of the glueball mass is defined by $\delta_G(L)
= 1-M(L)/\bar{M}$. The results for these fractional changes
are shown in Fig.~\ref{fvm}. Each point in the figure shows the
fractional change $\delta_G(L)$ of the glueball mass, the error
bars come from the statistical errors of $M(L)$, the solid lines
indicate $\delta_G=0$, the dotted lines above and below the solid
lines indicate $\delta_G = 0.02$ and $\delta_G = -0.02$, respectively. 
All changes are statistically consistent with zero, suggesting that
systematic errors in these results from finite volume are no
larger than the statistical errors for these physical volumes.
Since the physical volumes of the other $\beta$ values are not
smaller than the $8^3 \times 40$ at $\beta = 2.4$, we shall neglect
the FVE for the higher $\beta$ results.
\begin{table}
\caption{\label{fve_mass} The fitted ground state masses in the twenty
$R^{PC}$ channels at $\beta=2.4$ and $\xi=5$ on three different lattices
$L^3\times T = 16^3\times 80$, $12^3\times 64$, and $8^3\times 40$. }
\begin{ruledtabular} \begin{tabular}{cccc}
    & $L=16$    & $L=12$    & $L=8$ \\\hline
$A_1^{++}$ &0.308(3)    &0.308(2)    &0.312(2) \\
$A_1^{+-}$ &1.082(14)   &1.051(13)   &1.079(10)\\
$A_1^{-+}$ &0.604(11)   &0.618(4)    &0.618(3) \\
$A_1^{--}$ &1.03(3)     &1.075(16)   &1.01(3)  \\
       &         &            &         \\
$A_2^{++}$ &0.825(8)    &0.818(8)    &0.820(5) \\
$A_2^{+-}$ &0.805(6)    &0.804(8)    &0.808(5) \\
$A_2^{-+}$ &1.047(10)   &1.067(12)   &1.053(12)\\
$A_2^{--}$ &0.993(11)   &0.976(10)   &0.994(27)  \\
       &         &            &         \\
$E^{++} $  &0.542(3)    &0.536(2)    &0.541(2) \\
$E^{+-} $  &0.919(17)   &0.957(8)    &0.960(6) \\
$E^{-+} $  &0.699(4)    &0.698(4)    &0.695(3) \\
$E^{--} $  &0.879(5)    &0.891(7)    &0.882(5) \\
       &         &            &         \\
$T_1^{++}$ &0.826(5)    &0.832(4)    &0.834(3) \\
$T_1^{+-}$ &0.657(6)    &0.661(5)    &0.663(4) \\
$T_1^{-+}$ &0.932(5)    &0.940(5)    &0.936(4)\\
$T_1^{--}$ &0.865(12)   &0.884(5)    &0.893(4) \\
           &         &            &         \\
$T_2^{++}$ &0.542(3)    &0.536(3)    &0.538(2) \\
$T_2^{+-}$ &0.807(4)    &0.808(11)   &0.799(8)\\
$T_2^{-+}$ &0.697(5)    &0.700(4)    &0.700(2) \\
$T_2^{--}$ &0.903(4)    &0.893(6)    &0.896(3)
\end{tabular}
\end{ruledtabular}
\end{table}

\par
\begin{figure}
\includegraphics[height=7.5cm]{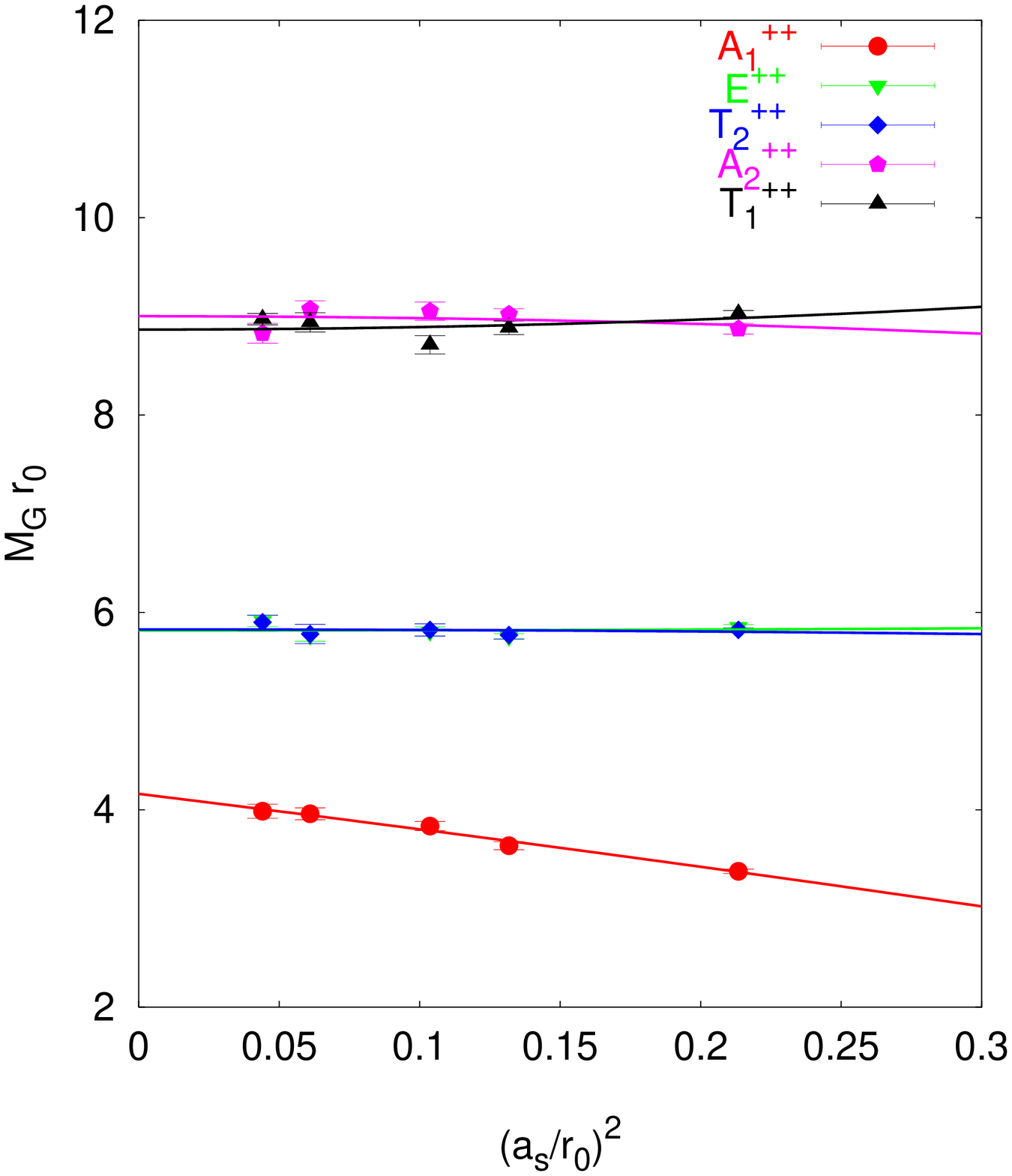}
\caption{\label{PC++}Masses of $PC=++$ glueballs in
terms of $r_0$ against the lattice spacing square $(a_s/r_0)^2$. The
calculated values are plotted in points, and the curves are the best
fits.}
\end{figure}
\begin{figure}
\includegraphics[height=7.5cm]{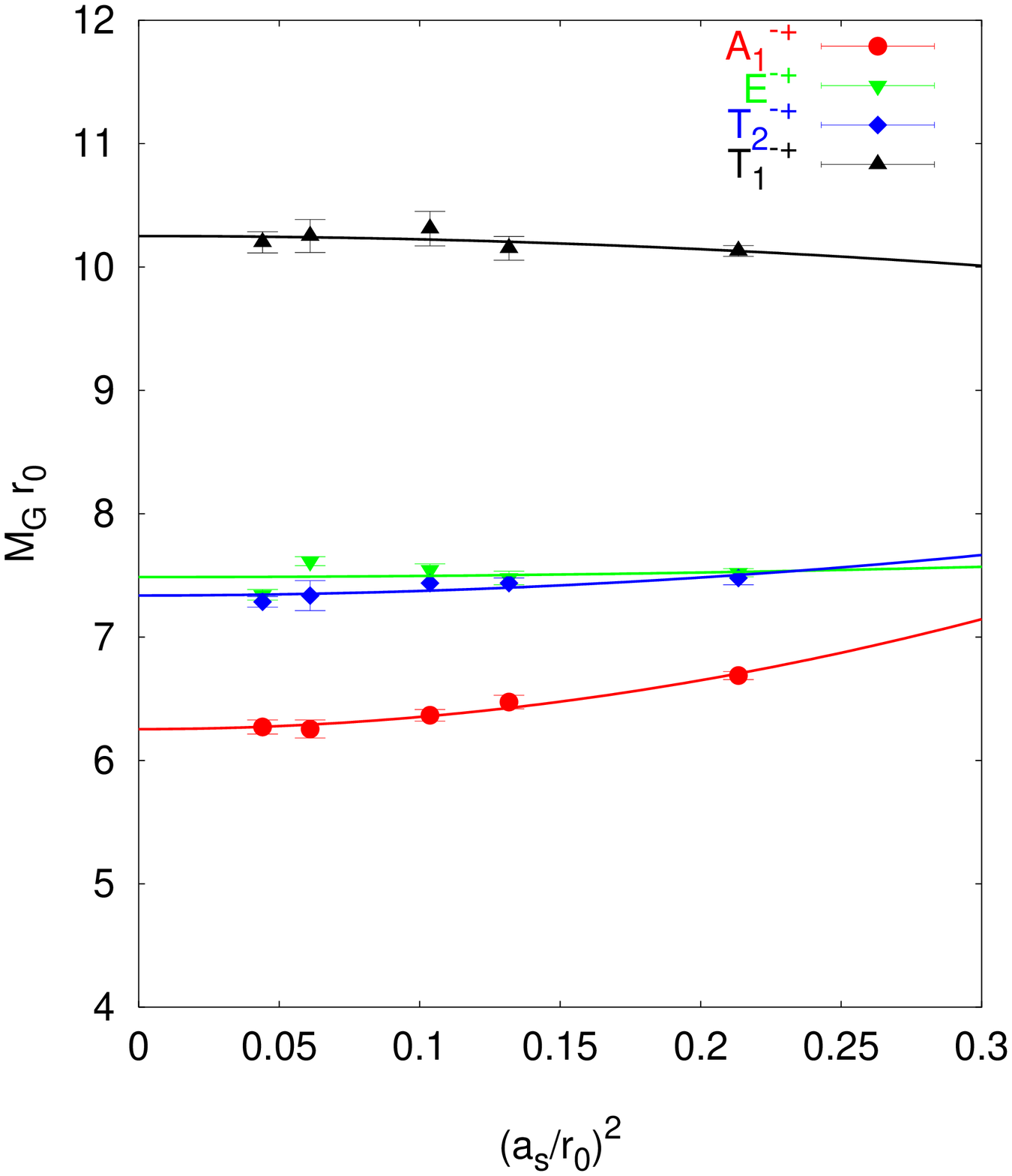}
\caption{\label{PC-+}Masses of $PC=-+$ glueballs in
terms of $r_0$ against the lattice spacing square $(a_s/r_0)^2$. The
calculated values are plotted in points, and the curves are the best
fits.}
\end{figure}
\begin{figure}
\includegraphics[height=7.5cm]{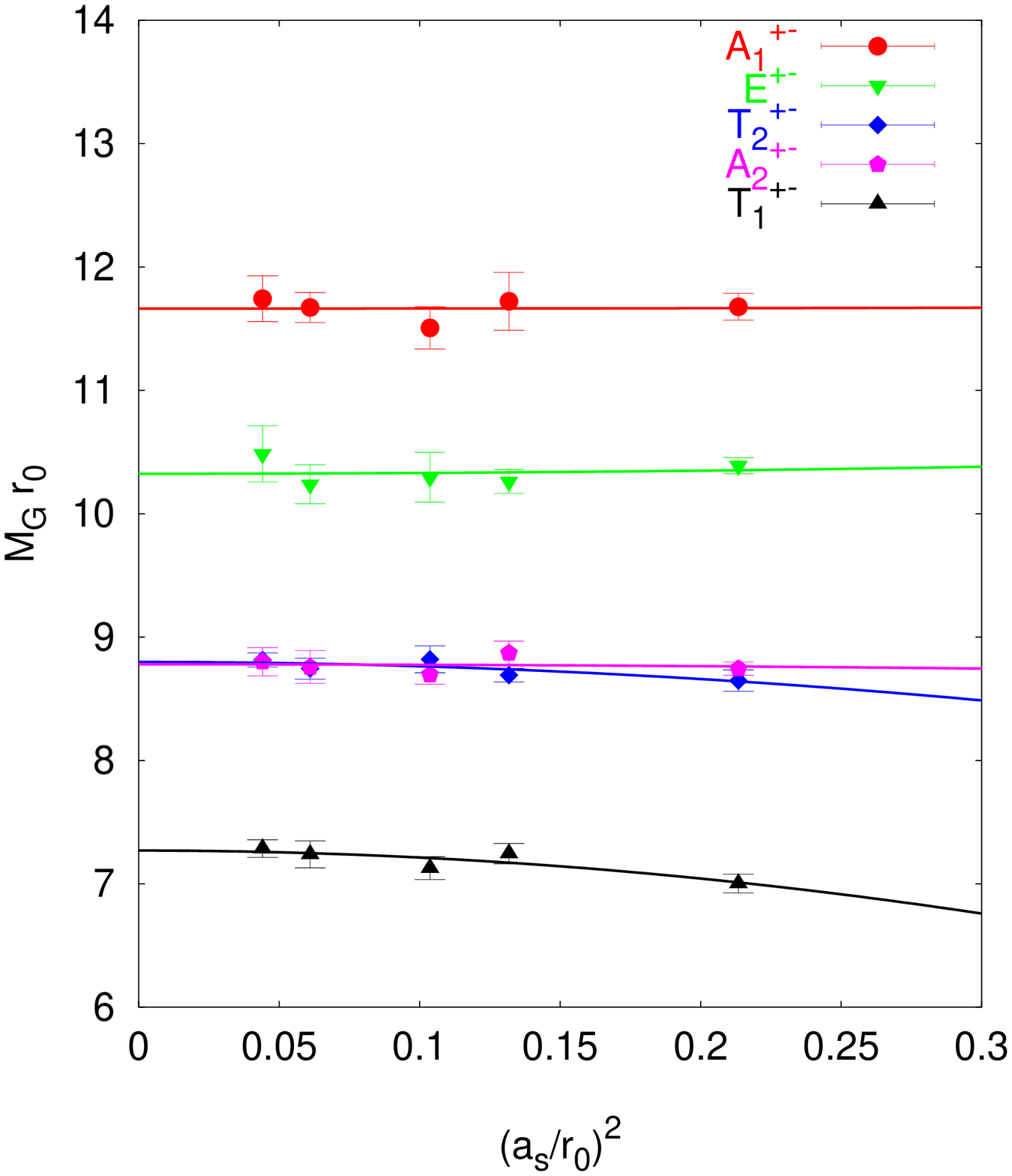}
\caption{\label{PC+-}Masses of $PC=+-$ glueballs in
terms of $r_0$ against the lattice spacing square $(a_s/r_0)^2$. The
calculated values are plotted in points, and the curves are the best
fits.}
\end{figure}
\begin{figure}
\includegraphics[height=7.5cm]{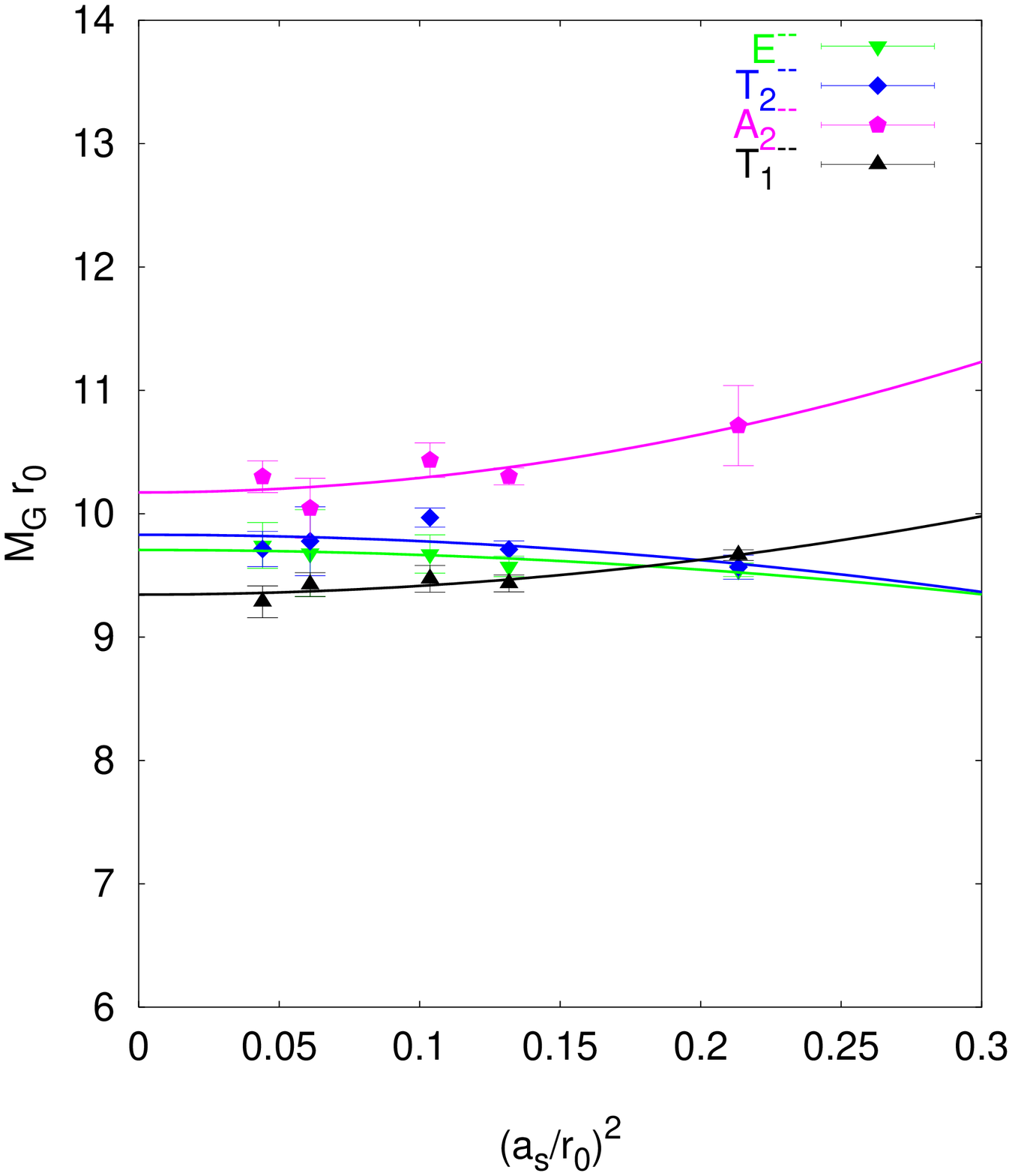}
\caption{\label{PC--}Masses of $PC=--$ glueballs in
terms of $r_0$ against the lattice spacing square $(a_s/r_0)^2$. The
calculated values are plotted in points, and the curves are the best
fits.}
\end{figure}

The fitted glueball masses at different coupling constant $\beta$ are
listed in Table~\ref{tab:masses}, where the statistical errors are
also quoted. The dimensionless products of $r_0$ and the glueball masses
$M_G$ are shown as functions of $(a_s/r_0)^2$. To remove
discretization errors from our estimates, the results for each level in
these figures must be extrapolated to the continuum limit
$a_s/r_0\rightarrow 0$. From perturbation theory, the leading
discretization errors are expected to be $O(a_t^2, a_s^4,
\alpha_s a_s^2)$. As discussed in Ref.~\cite{old3}, the $O(a_t^2,
\alpha_s a_s^2)$ errors could be negligible compared to the $O(a_s^4)$
errors for most calculated glueball masses except for $A_1^{++}$
glueball, so the fit model for these glueball masses is chosen to be
\begin{equation}
M_G(a_s)r_0 = M_G(0)r_0 +c_4 \frac{a_s^4}{r_0^4}.
\end{equation}
The masses versus $(a_s/r_0)^2$, as well as the fitted curves are
shown in Fig.~\ref{PC++},~\ref{PC-+},~\ref{PC+-}, and ~\ref{PC--}.
One can find from the figures that the data obey this function
very well. However, for $A_1^{++}$ glueball mass, the $O(a_s^2)$
error seems still very large. So we keep the linear term of
$a_s^2$ in the fit model, namely,
 \begin{equation}
M_G(a_s)r_0 = M_G(0)r_0 + c_2\frac{a_s^2}{r_0^2}+c_4 \frac{a_s^4}{r_0^4}.
\end{equation}
Already there has been some discussion on the possible reason for this 
large
discretization error in scalar channel; one can refer to
Ref.~\cite{old3} for details.
\par
We list several continuum extrapolated glueball masses in
Table~\ref{comparison}. We note that the earlier work~\cite{old3}
was carried out with $\beta = 1.7, 1.9, 2.2, 2,4, 2.5$ and 3.0. In
the present work, we concentrate on finer lattice spacings with
$\beta = 2.4, 2.6, 2.7, 3.0$ and 3.2. For comparison, we have
listed the results from the previous work~\cite{old3}.


\begin{table}
\caption{\label{tab:masses} The fitted ground state masses $M$ in the
twenty $R^{PC}$ channels at five different $\beta$'s. The errors
quoted here are all statistical errors from correlated
minimal-$\chi^2$ method.}
\begin{ruledtabular}
\begin{tabular}{llllll}
$R^{PC}$    & $\beta=2.4$ & $\beta=2.6$ & $\beta=2.7$ &
$\beta=3.0$ & $\beta=3.2$ \\ \hline

$A_1^{++}$ &0.312(2) &0.264(2) &0.247(2) &0.325(3) &0.279(5) \\
$A_1^{+-}$ &1.079(10)&0.851(17) &0.741(11)&0.961(10)&0.822(13)\\
$A_1^{-+}$ &0.618(3) &0.470(4) &0.410(3) &0.515(6) &0.439(4) \\
$A_1^{--}$ &1.01(3)  &0.897(20) &0.815(5) &1.036(12)&0.879(19)\\
       &         &         &     &         &         \\
$A_2^{++}$ &0.820(5) &0.665(4) &0.593(6) &0.779(17)&0.683(17)\\
$A_2^{+-}$ &0.808(5) &0.644(7) &0.560(5) &0.721(11)&0.616(8)\\
$A_2^{-+}$ &1.053(12)&0.850(7)&0.745(26)&1.02(4)&0.868(18)  \\
$A_2^{--}$ &0.994(27)&0.748(5) &0.672(9)&0.827(21)&0.721(9)\\
       &         &         &     &         &         \\
$E^{++} $  &0.541(2) &0.418(2) &0.374(3) &0.475(5) &0.414(4) \\
$E^{+-} $  &0.960(6) &0.745(7) &0.663(13) &0.843(13) &0.734(16)\\
$E^{-+} $  &0.695(3) &0.543(4) &0.486(3) &0.627(3) &0.514(3) \\
$E^{--} $  &0.882(5) &0.695(6) &0.623(10)&0.797(29)&0.682(13)\\
       &         &         &         &         &         \\
$T_1^{++}$ &0.834(3) &0.645(5) &0.561(3) &0.736(8) &0.628(4) \\
$T_1^{+-}$ &0.663(4) &0.526(6) &0.459(6) &0.596(9) &0.510(5)\\
$T_1^{-+}$ &0.936(4) &0.737(7) &0.664(9) &0.844(11)&0.714(6) \\
$T_1^{--}$ &0.893(4) &0.685(5) &0.610(7) &0.776(8) &0.650(9)\\
           &         &         &     &         &         \\
$T_2^{++}$ &0.538(2) &0.419(2) &0.375(4) &0.476(8) &0.413(5) \\
$T_2^{+-}$ &0.799(8) &0.631(4) &0.568(7) &0.720(7) &0.617(4) \\
$T_2^{-+}$ &0.700(2) &0.540(3) &0.479(4) &0.604(10)&0.510(3) \\
$T_2^{--}$ &0.896(3) &0.705(5) &0.642(5) &0.805(23)&0.680(10) \\
\end{tabular}
\end{ruledtabular}
\end{table}
\begin{table}
\begin{ruledtabular}
\caption{\label{comparison}  Continuum-limit glueball masses
$M_G$. The corresponding results in
Reference~\cite{old3} are also quoted for comparison.}
\begin{tabular}{cccc}
$R^{PC}$  &  Possible $J^{PC}$& $r_0 M_G$ (this work) & $r_0
M_G$~\cite{old3}\\
\hline
$A_1^{++}$&  $0^{++}$ & 4.16(11)   &    4.21(11)    \\

$E^{++}  $&  $2^{++}$ & 5.82(5)    &    5.85(2)     \\
$T_2^{++}$&  $2^{++}$ & 5.83(4)    &    5.85(2)     \\

$A_2^{++}$&  $3^{++}$ & 9.00(8)    &    8.99(4)     \\
$T_1^{++}$&  $3^{++}$ & 8.87(8)    &    8.99(4)     \\

$A_1^{-+}$&  $0^{-+}$ & 6.25(6)    &    6.33(7)     \\

$T_1^{+-}$&  $1^{+-}$ & 7.27(4)   &    7.18(3)     \\

$E^{-+}  $&  $2^{-+}$ & 7.49(7)    &    7.55(3)     \\
$T_2^{-+}$&  $2^{-+}$ & 7.34(11)   &    7.55(3)     \\

$T_2^{+-}$&  $3^{+-}$ & 8.80(3)   &     8.66(4)     \\
$A_2^{+-}$&  $3^{+-}$ & 8.78(5)   &    8.66(3)     \\

$T_1^{--}$&  $1^{--}$ & 9.34(4)   &    9.50(4)     \\

$E^{--}$&    $2^{--}$ & 9.71(3)   &    9.59(4)     \\
$T_2^{--}$&  $2^{--}$ & 9.83(8)   &    9.59(4)     \\

$A_2^{--}$&  $3^{--}$ & 10.25(4)   &    10.06(21)     \\

$E^{+-}$&    $2^{+-}$ & 10.32(7)   &    10.10(7)     \\

$A_1^{+-}$&  $0^{+-}$ & 11.66(7)   &    11.57(12)     \\
\end{tabular}
\end{ruledtabular}
\end{table}

\subsection*{C. The Glueball Matrix Elements}
\begin{figure}[th]
\includegraphics[height=7.5cm]{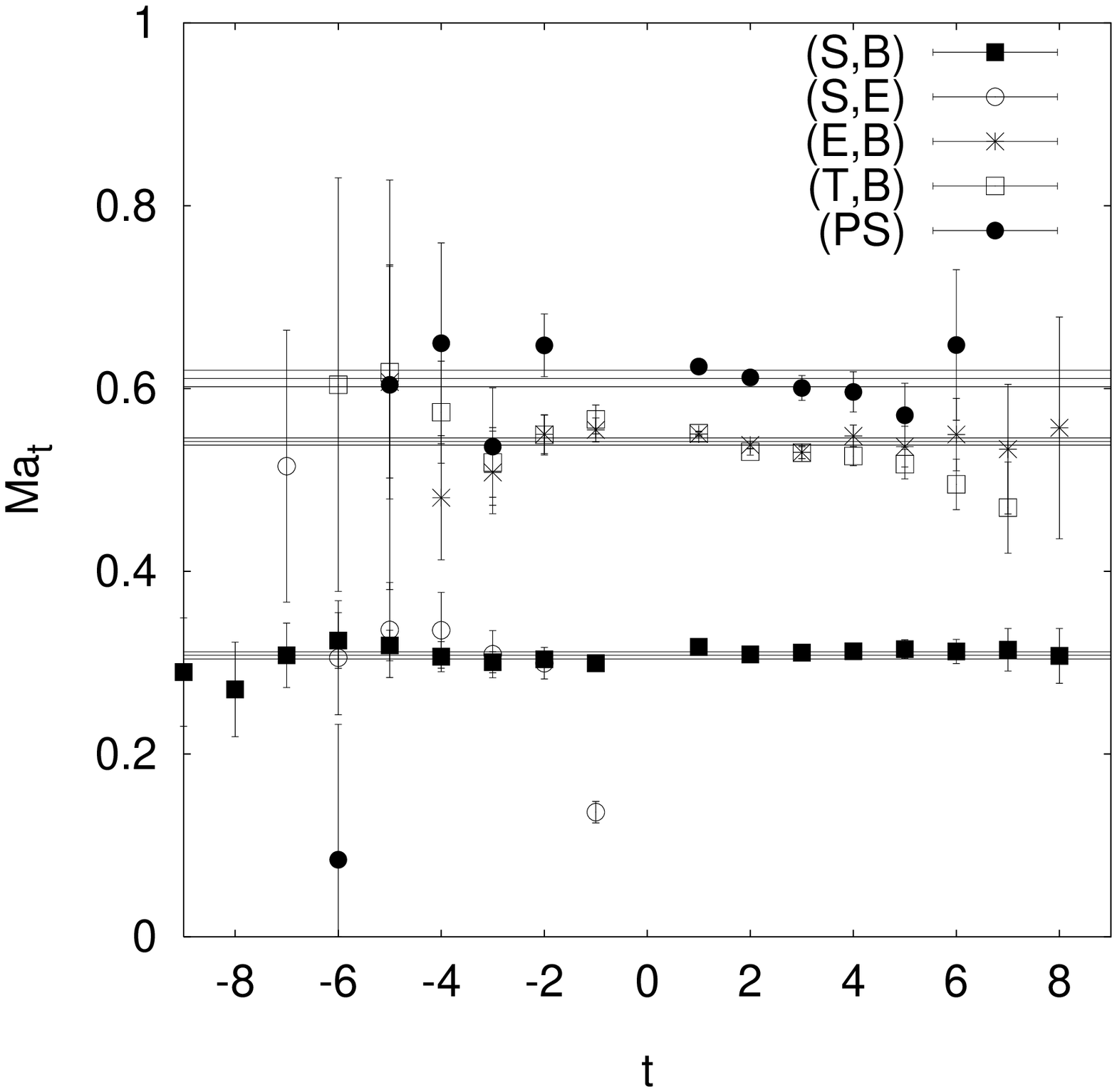}
\caption{\label{mtxfit1} The effective mass plateaus of the
smeared-smeared and smeared-local correlation functions at
$\beta=2.4, \xi=5$ on the lattice $8^3\times 40$. The
smeared-local correlation functions are plotted on the
negative-time side of the $t$ axis. The Type-I local
operators are used.}
\end{figure}
\begin{figure}[th]
\includegraphics[height=7.5cm]{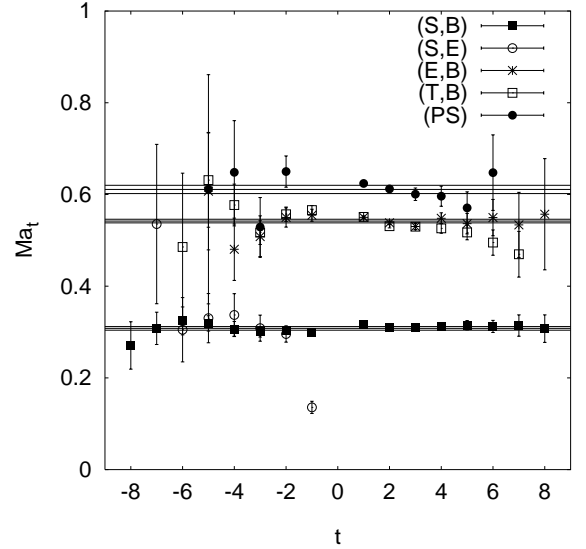}
\caption{\label{mtxfit2} The effective mass plateaus of the
smeared-smeared and smeared-local correlation functions at
$\beta=2.4, \xi=5$ on the lattice $8^3\times 40$. The
smeared-local correlation functions are plotted on the
negative-time side of the $t$ axis. The Type-II local operators
are used.}
\end{figure}

As described above, we use the correlated minimal-$\chi^2$
method to fit the smeared-smeared and smeared-local two-point functions,
namely, $C_{SS}(t)$ and $C_{SL}(t)$, simultaneously using the fit
functions
\begin{eqnarray}
C_{SS}(t) &=& X^2\left(e^{-M t}+e^{-M(T-t)}\right) \nonumber\\
C_{SL}(t) &=& X Y\left(e^{-M t}+e^{-M(T-t)}\right),
\end{eqnarray}
where $X=\langle 0|O_S(0)|G\rangle/\sqrt{2M_G V}$ and $Y=\langle
0|O_L(0)|G\rangle/\sqrt{2M_G V}$. Here $V$ is the spatial lattice volume 
in physical units, say, $V=L^3a_s^3$, and
$M_G=M a_t^{-1}$. As an example, we show 
the effective mass plot at $\beta=2.4$ ($L=8$ lattice) in
Figs.~\ref{mtxfit1},
and ~\ref{mtxfit2}, 
where the smeared-local correlation functions are plotted on the
negative-time side of the $t$ axis. In each channel, the
effective mass plateau for $C_{SS}$ are flatter than that of
$C_{SL}$. Nevertheless, both of them can be fitted by the same
mass parameter consistently. For all the five $\beta$'s, the fit
parameters $X$,$Y$, effective masses $M$, as well as the fit
windows, are listed in Table~\ref{table24}, \ref{table26},
\ref{table27}, \ref{table30}, and \ref{table32}. In these tables,
the first group of data is for Type-I operators, the second group
for Type-II operators, and the last two row of data for the second
definition (Eq.~(\ref{imag2})) of Type-II operators in $T_2$ irreps. We 
give a brief
interpretation of the meaning of the value of $X$ parameter here.
During the practical performance of the variational method
discussed in Sec. III,  the eigenvectors ${\bf v}^{(R)}$ in
Eq.~(\ref{eigen}) are normalized as $ {{\bf v}^{(R)}}^T
\tilde{C}(0){\bf v}^{(R)} = 1$, so that the parameter $X$
indicates the relative overlap of the smeared operator to the
specific stationary state (the ground state here). In these
tables, all the $X$ parameters are very close to 1 which implies
that the smeared operators couple almost completely to the ground
state and there is little contamination from excited states. The
$\chi^2$'s per degree of freedom of the data fitting are all
acceptable and  also listed in the tables.

\begin{table*}
\caption{ \label{table24} The matrix elements extracted at $\beta
= 2.4$ (on lattice $8^3\times 40$). The smeared-smeared(SS) and 
smeared-local(SL) correlators
are fitted simultaneously in the time window $t$(SS) and $t$(SL). The
fitted masses $M$, $X$, and $Y$ parameters are listed. The first group 
of data is for Type-I operators, the second group
for Type-II operators, and the last two row of data for the second
definition (Eq.~(\ref{imag2})) of Type-II operators in $T_2$ irreps.}

\begin{ruledtabular}
\begin{tabular}{ccccccc}
  R  &   $t$(SS) & $t$(SL) &  $M$   &$Y$(${}\times 10^{-2}$)&$X$&
$\chi^2/d.o.f$\\
\hline
  $(S,B)$ &  1 -- 4 & 1 -- 3 & 0.311(2) & 21.3(1)   &   0.994(3)  & 0.92 
\\
  $(S,E)$ &  1 -- 4 & 1 -- 4 & 0.311(2) & 21.1(2)   &   0.994(3)  & 0.69 
\\
  $(PS)$ &  1 -- 5 & 2 -- 4 & 0.617(4) & 5.24(11)  &   0.985(4)  & 1.59 
\\
  $(E,B)$ &  1 -- 7 & 1 -- 5 & 0.541(2) & 2.64(3)   &   0.995(2)  & 0.64 
\\
  $(E,E)$ &  1 -- 7 & 1 -- 4 & 0.541(2) & 1.52(9)   &   0.995(2)  & 0.64 
\\
  $(T_2,B)$ &  2 -- 7 & 1 -- 4 & 0.539(3) & 1.84(12)  &   0.989(2)  & 
0.57 \\
  $(T_2,E)$ &  2 -- 7 & 1 -- 3 & 0.539(3) & 0.85(5)   &   0.990(2)  & 
0.52 \\
\hline
  $(S,B)$ &  1 -- 9 & 1 -- 7 & 0.311(2) & 11.54(6)  &   0.993(3)  & 1.12 
\\
  $(S,E)$ &  1 -- 9 & 1 -- 7 & 0.311(2) & 16.38(13) &   0.995(3)  & 0.75 
\\
  $(PS)$ &  2 -- 7 & 1 -- 6 & 0.617(4) & 5.33(11)  &   0.985(4)  & 1.36 
\\
  $(E,B)$ &  1 -- 7 & 1 -- 6 & 0.541(2) & 2.50(2)   &   0.995(2)  & 
0.89 \\
  $(E,E)$ &  1 -- 7 & 1 -- 3 & 0.542(2) & 1.35(4)   &   0.995(2)  & 0.87 
\\
  $(T_2,B)$ &  2 -- 7 & 1 -- 5 & 0.541(2) & 2.24(1)   &   0.990(3)  & 
1.34 \\
  $(T_2,E)$ &  2 -- 7 & 2 -- 4 & 0.542(2) & 0.73(4)   &   0.989(2)  & 
0.45 \\
\hline
  $(T_2,B)$ &  2 -- 7 & 2 -- 5 & 0.543(2) & 2.37(1)  &   0.990(2)  & 
1.06 \\
  $(T_2,E)$ &  2 -- 7 & 1 -- 4 & 0.542(2) & 0.73(4) &   0.989(2)  & 
0.46 \\

\end{tabular}
\end{ruledtabular}
\end{table*}

\begin{table*}
\caption{ \label{table26}
The same as in Table~\ref{table24}, but for $\beta = 2.6$}
\begin{ruledtabular}
\begin{tabular}{ccccccc}
  R  &   $t$(SS) & $t$(SL) &  $M$   &$Y$(${}\times 10^{-2}$)&$X$&
$\chi^2/d.o.f$\\
\hline
  $(S,B)$ &  1 -- 9 & 1 -- 7 & 0.264(2) & 6.48(5)   &   0.994(3)  & 0.92 
\\
  $(S,E)$ &  1 -- 9 & 1 -- 7 & 0.264(2) & 6.37(11)  &   0.994(3)  & 0.69 
\\
  $(PS)$ &  2 -- 7 & 1 -- 5 & 0.470(4) & 2.14(3)   &   0.985(4)  & 1.59 
\\
  $(E,B)$ &  1 -- 7 & 1 -- 6 & 0.424(2) & 0.716(10) &   0.995(2)  & 0.64 
\\
  $(E,E)$ &  1 -- 7 & 1 -- 3 & 0.423(2) & 0.39(4)   &   0.995(2)  & 0.64 
\\
  $(T_2,B)$ &  2 -- 7 & 1 -- 5 & 0.421(1) & 0.562(7)  &   0.989(2)  & 
0.57 \\
  $(T_2,E)$ &  2 -- 7 & 1 -- 4 & 0.421(2) & 0.23(2)   &   0.990(2)  & 
0.52 \\
\hline
  $(S,B)$ &  1 -- 9 & 1 -- 7 & 0.265(2) & 4.01(3)   &   0.993(3)  & 1.12 
\\
  $(S,E)$ &  1 -- 9 & 1 -- 7 & 0.263(2) & 5.06(3)   &   0.995(3)  & 0.75 
\\
  $(PS)$ &  2 -- 7 & 1 -- 6 & 0.470(5) & 2.16(3)   &   0.985(4)  & 1.36 
\\
  $(E,B)$ &  1 -- 7 & 1 -- 6 & 0.424(2) & 0.706(13) &   0.995(2)  & 0.89 
\\
  $(E,E)$ &  1 -- 7 & 1 -- 3 & 0.424(2) & 0.40(3)   &   0.995(2)  & 0.87 
\\
  $(T_2,B)$ &  2 -- 7 & 1 -- 5 & 0.423(2) & 0.698(5)  &   0.990(3)  & 
1.34 \\
  $(T_2,E)$ &  2 -- 7 & 2 -- 4 & 0.421(2) & 0.22(2)   &   0.989(2)  & 
0.45 \\
\hline
  $(T_2,B)$ &  2 -- 7 & 2 -- 5 & 0.422(2) & 0.719(10) &   0.990(2)  & 
1.06 \\
  $(T_2,E)$ &  2 -- 7 & 1 -- 4 & 0.421(2) & 0.22(2)   &   0.989(2)  & 
0.46 \\
\end{tabular}
\end{ruledtabular}
\end{table*}

\begin{table*}
\caption{ \label{table27}
The same as in Table~\ref{table24}, but for $\beta = 2.7$}
\begin{ruledtabular}
\begin{tabular}{ccccccc}
  R  &   $t$(SS) & $t$(SL) &  $M$   &$Y$(${}\times 10^{-2}$)&$X$&
$\chi^2/d.o.f$\\
\hline
  $(S,B)$  &  2 -- 9 &  1 -- 7 & 0.247(2) &  5.03(4)   &  0.990(3) & 
0.92 \\
  $(S,E)$  &  2 -- 9 &  1 -- 7 & 0.246(2) &  4.88(13)  &  0.991(3) & 
0.50 \\
  $(PS)$  &  1 -- 7 &  2 -- 6 & 0.431(3) &  1.88(5)   &  0.991(2) & 1.07 
\\
  $(E,B)$  &  3 -- 8 &  1 -- 4 & 0.376(3) &  0.575(10) &  0.982(4) & 
0.75 \\
  $(E,E)$  &  3 -- 8 &  1 -- 3 & 0.376(4) &  0.27(3)   &  0.981(4) & 
0.39 \\
  $(T_2,B)$  &  4 -- 8 &  2 -- 6 & 0.379(5) &  0.423(10) &  0.977(7) & 
1.14 \\
  $(T_2,E)$  &  4 -- 8 &  1 -- 4 & 0.378(6) &  0.20(2)   &  0.975(8) & 
1.39 \\
\hline
  $(S,B)$  &  2 -- 9 &  3 -- 7 & 0.246(2) &  3.21(4)   &  0.990(3) & 
0.86 \\
  $(S,E)$  &  2 -- 9 &  3 -- 7 & 0.247(2) &  3.80(10)  &  0.990(3) & 
0.58 \\
  $(PS)$  &  2 -- 7 &  1 -- 6 & 0.431(2) &  1.90(5)   &  0.991(2) & 1.08 
\\
  $(E,B)$  &  3 -- 8 &  2 -- 5 & 0.376(3) &  0.592(10) &  0.981(4) & 
0.60 \\
  $(E,E)$  &  3 -- 8 &  2 -- 5 & 0.376(3) &  0.27(2)   &  0.982(4) & 
0.22 \\
  $(T_2,B)$  &  4 -- 8 &  2 -- 6 & 0.386(5) &  0.546(10) &  0.988(8) & 
1.79 \\
  $(T_2,E)$  &  4 -- 8 &  1 -- 3 & 0.378(5) &  0.19(2)   &  0.975(8) & 
1.28 \\
\hline
  $(T_2,B)$  &  4 -- 8 &  3 -- 6 & 0.386(5) &  0.565(9)  &  0.987(8) & 
1.82 \\
  $(T_2,E)$  &  4 -- 8 &  2 -- 3 & 0.378(5) &  0.19(2)   &  0.975(8) & 
1.29 \\
\end{tabular}
\end{ruledtabular}
\end{table*}

\begin{table*}
\caption{ \label{table30}
The same as in Table~\ref{table24}, but for $\beta = 3.0$}
\begin{ruledtabular}
\begin{tabular}{ccccccc}
  R  &   $t$(SS) & $t$(SL) &  $M$   &$Y$(${}\times 10^{-2}$)&$X$&
$\chi^2/d.o.f$\\
\hline
  $(S,B)$ &  2 -- 9 &  1 -- 4 & 0.324(3) &  1.56(3)    &  0.985(3) & 
0.99\\
  $(S,E)$ &  2 -- 9 &  1 -- 4 & 0.326(4) &  1.50(4)    &  0.985(3) & 
0.82\\
  $(PS)$ &  2 -- 5 &  1 -- 4 & 0.542(6) &  0.706(15)  &  0.981(6) & 
2.00\\
  $(E,B)$ &  3 -- 7 &  1 -- 4 & 0.480(6) &  0.176(7)   &  0.967(8) & 
0.22\\
  $(E,E)$ &  3 -- 7 &  1 -- 3 & 0.480(6) &  0.107(14)  &  0.967(8) & 
0.25\\
  $(T_2,B)$ &  3 -- 8 &  1 -- 3 & 0.487(5) &  0.137(5)   &  0.969(7) & 
1.28\\
  $(T_2,E)$ &  3 -- 8 &  1 -- 3 & 0.485(6) &  0.060(9)   &  0.967(7) & 
0.94\\
\hline
  $(S,B)$ &  2 -- 9 &  2 -- 7 & 0.326(4) &  1.07(2)    &  0.984(3) & 
0.68\\
  $(S,E)$ &  2 -- 9 &  1 -- 8 & 0.324(3) &  1.09(3)    &  0.986(3) & 
0.49\\
  $(PS)$ &  2 -- 5 &  1 -- 4 & 0.541(6) &  0.715(15)  &  0.981(5) & 
1.84\\
  $(E,B)$ &  3 -- 7 &  1 -- 4 & 0.477(6) &  0.179(5)   &  0.964(8) & 
0.72\\
  $(E,E)$ &  3 -- 7 &  1 -- 3 & 0.480(6) &  0.086(5)   &  0.967(8) & 
0.22\\
  $(T_2,B)$ &  3 -- 8 &  1 -- 4 & 0.490(5) &  0.172(4)   &  0.971(7) & 
1.55\\
  $(T_2,E)$ &  3 -- 8 &  1 -- 3 & 0.485(6) &  0.053(6)   &  0.968(8) & 
1.17\\
\hline
  $(T_2,B)$ &  3 -- 8 &  1 -- 4 & 0.489(5) &  0.179(4)   &  0.971(7) & 
1.46\\
  $(T_2,E)$ &  3 -- 8 &  1 -- 3 & 0.485(6) &  0.053(6)   &  0.968(8) & 
1.17\\
\end{tabular}
\end{ruledtabular}

\end{table*}

\begin{table*}
\caption{ \label{table32}
The same as in Table~\ref{table24}, but for $\beta = 3.2$}
\begin{ruledtabular}
\begin{tabular}{ccccccc}
  R  &   $t$(SS) & $t$(SL) &  $M$   &$Y$(${}\times 10^{-2}$)&$X$&
$\chi^2/d.o.f$\\
\hline
  $(S,B)$ & 4 --  7 &  1 -- 6 & 0.279(6) & 0.518(11)   & 0.966(9) & 
1.74\\
  $(S,E)$ & 4 --  7 &  1 -- 4 & 0.278(5) & 0.534(26)   & 0.965(9) & 
0.18\\
  $(PS)$ & 3 --  9 &  3 -- 6 & 0.439(8) & 0.294(15)   & 0.932(9) & 
0.38\\
  $(E,B)$ & 3 --  7 &  1 -- 4 & 0.417(4) & 0.062(4)    & 0.967(3) & 
2.77\\
  $(E,E)$ & 3 --  7 &  1 -- 3 & 0.415(4) & 0.048(8)    & 0.965(3) & 
1.49\\
  $(T_2,B)$ & 3 --  9 &  1 -- 4 & 0.414(3) & 0.0548(13)  & 0.959(4) & 
0.67\\
  $(T_2,E)$ & 3 --  9 &  1 -- 3 & 0.415(3) & 0.033(5)    & 0.960(5) & 
0.67\\
\hline
  $(S,B)$ & 4 --  7 &  1 -- 8 & 0.282(5) & 0.391(7)    & 0.968(7) & 
1.98\\
  $(S,E)$ & 4 --  7 &  1 -- 6 & 0.280(5) & 0.376(11)   & 0.967(8) & 
0.31\\
  $(PS)$ & 3 --  9 &  3 -- 6 & 0.439(8) & 0.299(15)   & 0.932(9) & 
0.40\\
  $(E,B)$ & 3 --  7 &  1 -- 3 & 0.415(4) & 0.0693(23)  & 0.966(4) & 
3.29\\
  $(E,E)$ & 3 --  7 &  1 -- 3 & 0.416(4) & 0.037(4)    & 0.965(4) & 
3.05\\
  $(T_2,B)$ & 3 --  9 &  1 -- 7 & 0.416(4) & 0.068(2)    & 0.961(6) & 
0.96\\
  $(T_2,E)$ & 3 --  9 &  1 -- 3 & 0.415(3) & 0.028(3)    & 0.960(4) & 
0.56\\
\hline
  $(T_2,B)$ & 3 -  9 &  2 - 5 & 0.415(4) & 0.068(3)    & 0.960(5) & 
0.74\\
  $(T_2,B)$ & 3 -  9 &  1 - 3 & 0.415(3) & 0.028(3)    & 0.960(4) & 
0.54\\
\end{tabular}
\end{ruledtabular}
\end{table*}
\par
\begin{figure}
\includegraphics[height=6.5cm]{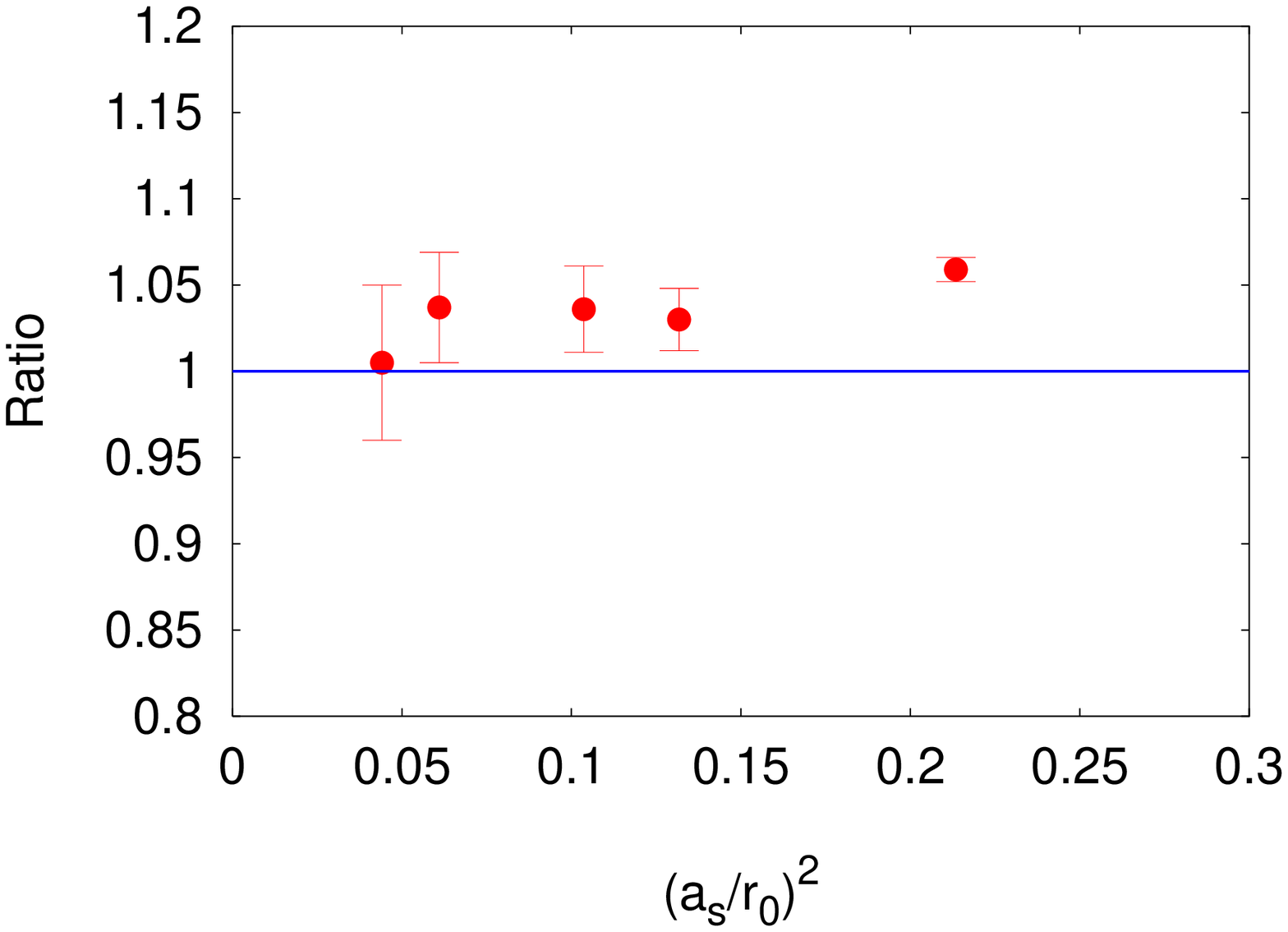}
\caption[]{\label{artifacts} The data points represent the ratios of 
the matrix elements of the two definitions of operator $(T_2,B)$ at 
different lattice spacings. The straight line is plotted to show how far 
the ratios deviate from 1.  
}
\end{figure}

From the measured matrix elements of the two definitions of the Type-II
local operators in tensor channel ($T_2$ irreps), we can get an estimate
of the $O(a_s^4)$ lattice artifacts. Recalling the discussion in Sec. 
II,
after restoring the fact $ig$ to $F_{\mu\nu}$,
the difference of the two definition of the operators $(T_2,B)$ is
\begin{equation}
\Delta = \frac{1}{36}a_s^8 g^2(D_{\mu}^2+D_{\nu}^2)F_{\mu\nu}
                         (D_{\rho}^2+D_{\sigma}^2)F_{\rho\sigma} +
\ldots,
\end{equation}
with the indices $\mu$, $\nu$, $\rho$, and $\sigma$ varying accordingly.
The ratio of the matrix elements of the two
definitions of operator $(T_2,B)$ is plotted in Fig.~\ref{artifacts}
with respect to the lattice spacing. It is known that this difference
comes totally from the lattice discretization and will disappear in the
continuum limit. This is the case from the figure. Even though the
relative difference is less than 3--4\%, the deviation of the
ratio from 1 is detectable on coarse lattice with larger lattice
spacing and decreases when approaching to the continuum limit. On the
finest lattice we are using, the difference is consistent with zero
within the error. This result shows that, as far as the Type-II
operators are concerned, after the implementation of Symanzik's
improvement scheme along with the tadpole improvement, the measured
matrix elements of the two versions of lattice local operator have
a few percent differences at finite $a_s$, but approaches the same
continuum limit as $a_s$ goes to zero.

\par
\begin{figure}[htb!]
\includegraphics[height=7.5cm]{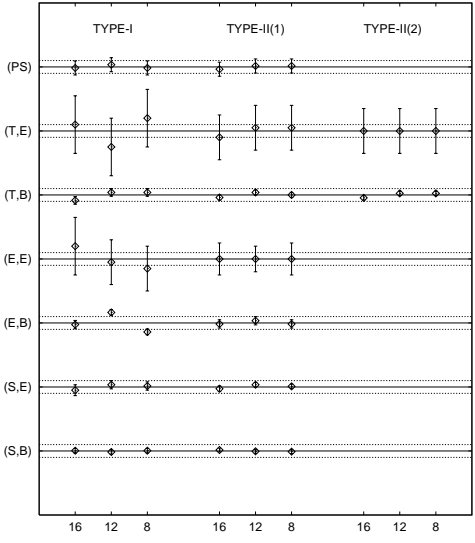}
\caption[]{\label{fve}
Finite-volume effects on the results of the $\beta=2.4$,
$\xi=5$ calculation. Each point shows the fractional change $\delta_G(L) =
1-f_G(L)/\bar{f}_G$ in the matrix element of a glueball $G$, where
$f_G(L)$ is the matrix element of $G$ measured on the lattice $L^3\times
T$ with $L = 8,12,$ and 16, and $\bar{f}_G$ is the average values over
those from different lattices. The errorbars come from the statistical
errors of $f_G(L)$. The lattice labels $L$ are shown along the
horizontal axis, and the labels of different local operators are
specified along the vertical axis. The solid lines indicates
$\delta_G=0$, and the dotted lines above the solid lines
indicates $\delta_G = 0.02$, and the dotted line lines below the solid
lines indicate $\delta_G = -0.02$. The FVE of different definitions of
local operators are also shown in the figure.}
\end{figure}

Here comes the discussion of the FVE of the matrix elements. By analogy 
with the FVE analysis of glueball masses, the glueball matrix elements 
are also calculated
at $\beta = 2.4$ and $\xi=5$ on a $8^3\times 40$ lattice, a
$12^3\times 64$, and a $16^3\times 80$ lattice. For these three
runs, all the input parameters are the same except the different
lattice volume. The extracted matrix elements are listed in
Table~\ref{fve_mtx1}, \ref{fve_mtx2}, and \ref{fve_mtx3},
respectively for Type-I and Type-II operators. As illustrated in
the tables, the finite volume effects of matrix elements are very
small and in most case the changes are consistent with zero
statistically. We also define the relative deviation $\delta_G(L)
= 1-f_G(L)/\bar{f}_G$ to show the FVE quantitatively, where
$\bar{f}_G$ is the mean value of the matrix elements of glueball
$G$ averaged over the three lattice volumes, $f_G(L)$ denote the
matrix element of glueball $G$ measured on lattice $L^3\times T$.
The results of $\delta_G(L)$ are shown in Fig.~\ref{fve}. In the
figure, each point shows $\delta_G(L)$ of the matrix element of a
glueball $G$, and the error bars come from the statistical errors
of $f_G(L)$. The solid lines indicates $\delta_G=0$, the dotted
lines above and below the solid lines indicate $\delta_G = \pm
0.02$. The largest effects from the finite volume appear to occur
in the TYPE I operator of $(E,B)$. All deviations are
statistically consistent with zero, suggesting that systematic
errors in these results from finite volume are no larger than the
statistical errors. Since the physical volumes of the other
$\beta$ values are not smaller than the $8^3 \times 40$ at $\beta =
2.4$, we shall neglect the finite volume effects for the higher
$\beta$ results.

\begin{table}
\caption{\label{fve_mtx1} The finite volume effects of matrix
elements at $\beta=2.4$ and $\xi=5$ on three different lattices
$L^3\times T = 16^3\times 80$, $12^3\times 64$, and $8^3\times
40$. The fitted $Y$ parameters are listed in normal numbers. For
clarity, the $Y$ parameters for lattice $L=12$ and $L=8$ are
re-scaled to $L=16$ due to the relation $Y=\langle
0|O_L(0)|G\rangle/\sqrt{2M_GV}$ and illustrated by bold numbers.
This table lists the $Y$ parameters of Type-I operators.}
\begin{ruledtabular}
\begin{tabular}{c|ccc}
     & $L=16$           & $L=12$           & $L=8$   \\\hline
         &                  &                  &         \\
$(S,B)$  &{\bf 7.53(5)}     & {\bf 7.50(5)}    &{\bf 7.52(5)}  \\
     &7.53(5)           &11.54(7)           &21.27(15)  \\
     &                  &                  &         \\
$(S,E)$  &{\bf 7.38(12)}     &{\bf 7.51(11)}     &{\bf 7.47(8)}  \\
     &7.38(12)           &11.56(17)           &21.12(23) \\
         &                  &                  &         \\
$(E,B)$  &{\bf 0.956(12)}   &{\bf 0.993(8) }   &{\bf 0.936(11)} \\
         &0.956(12)         &1.53(12)           &2.64(3)  \\
         &                  &                  &         \\
$(E,E)$  &{\bf 0.56(5)}     &{\bf 0.52(4)}     &{\bf 0.54(3)}  \\
     &0.56(5)           &0.81(6)           &1.52(9)  \\
         &                  &                  &         \\
$(T_2,B)$&{\bf 0.633(8)}     &{\bf 0.649(8)}     &{\bf 0.651(4)}  \\
     &0.633(8)           &1.000(12)           &1.841(11)  \\
         &                  &                  &         \\
$(T_2,E)$&{\bf 0.29(2)}     &{\bf 0.27(3)}     &{\bf 0.30(2)}  \\
     &0.29(2)           &0.42(4)           &0.85(6)  \\
         &                  &                  &         \\
$(PS)$   &{\bf 1.85(4)}     &{\bf 1.88(4)}      &{\bf 1.85(4)}   \\
         &1.85(4)        &2.89(6)           &5.24(11)  \\
\end{tabular}
\end{ruledtabular}
\end{table}
\begin{table}
\caption{ \label{fve_mtx2}The same as Table~\ref{fve_mtx1}, but
for Type-II operators.}
\begin{ruledtabular}
\begin{tabular}{c|ccc}
      & $L=16$          & $L=12$        & $L=8$ \\
\hline
      &                 &                   &         \\
$(S,B)$   &{\bf 0.410(3)}       &{\bf 0.408(3)}     &{\bf 0.408(2)} \\
      &4.10(3)     &6.29(4)           &11.54(6)  \\
          &                 &                   &         \\
$(S,E)$   &{\bf 5.73(6)}       &{\bf 5.82(4)}     &{\bf 0.579(4)} \\
      &5.73(6)         &8.97(6)        &16.4(1)  \\
          &                 &                   &         \\
$(E,B)$   &{\bf 0.885(11)}       &{\bf 0.895(9)}     &{\bf 0.883(7)} \\
      &0.885(11)     &1.378(14)           &2.50(2) \\
          &                 &                   &         \\
$(E,E)$   &{\bf 47(2)}       &{\bf 0.47(2)}     &{\bf 0.48(2)} \\
      &0.47(2)         &0.72(3)           &1.35(4)\\
          &                 &                   &         \\
$(T_2,B)$ &{\bf 0.783(5)}       &{\bf 0.792(4)}     &{\bf 0.790(4)} \\
      &0.783(5)         &1.220(6)           &2.24(1) \\
          &                 &                   &         \\
$(T_2,E)$ &{\bf 0.25(2)}      &{\bf 0.26(2)}    &{\bf 0.26(2)}\\
      &0.25(2)        &0.40(3)          &0.73(4)\\
          &                 &                   &         \\
$(PS)$    &{\bf 1.87(4)}       &{\bf 1.90(4)}     &{\bf 1.89(4)} \\
      &1.87(4)         &2.92(6)           &5.33(11) \\
\end{tabular}
\end{ruledtabular}
\end{table}

\begin{table}
\caption{\label{fve_mtx3}The same as Table~\ref{fve_mtx1}, but for
the second definition of Type-II operators $O(T_2,B)$ and
$O(T_2,E)$. }
\begin{ruledtabular}
\begin{tabular}{c|ccc}
      & $L=16$          & $L=12$        & $L=8$ \\
\hline
      &                 &                   &         \\
$(T_2,B)$ &{\bf 0.825(6)}       &{\bf 0.835(8)}     &{\bf 0.837(4)} \\
      &0.825(6)         &1.285(12)           &2.367(12) \\
          &                 &                   &         \\
$(T_2,E)$ &{\bf 0.26(2)}      &{\bf 0.26(2)}    &{\bf 0.26(2)}\\
      &0.26(2)        &0.40(3)          &0.73(4)\\
\end{tabular}
\end{ruledtabular}

\end{table}

\begin{figure*}
\includegraphics[height=7.5cm]{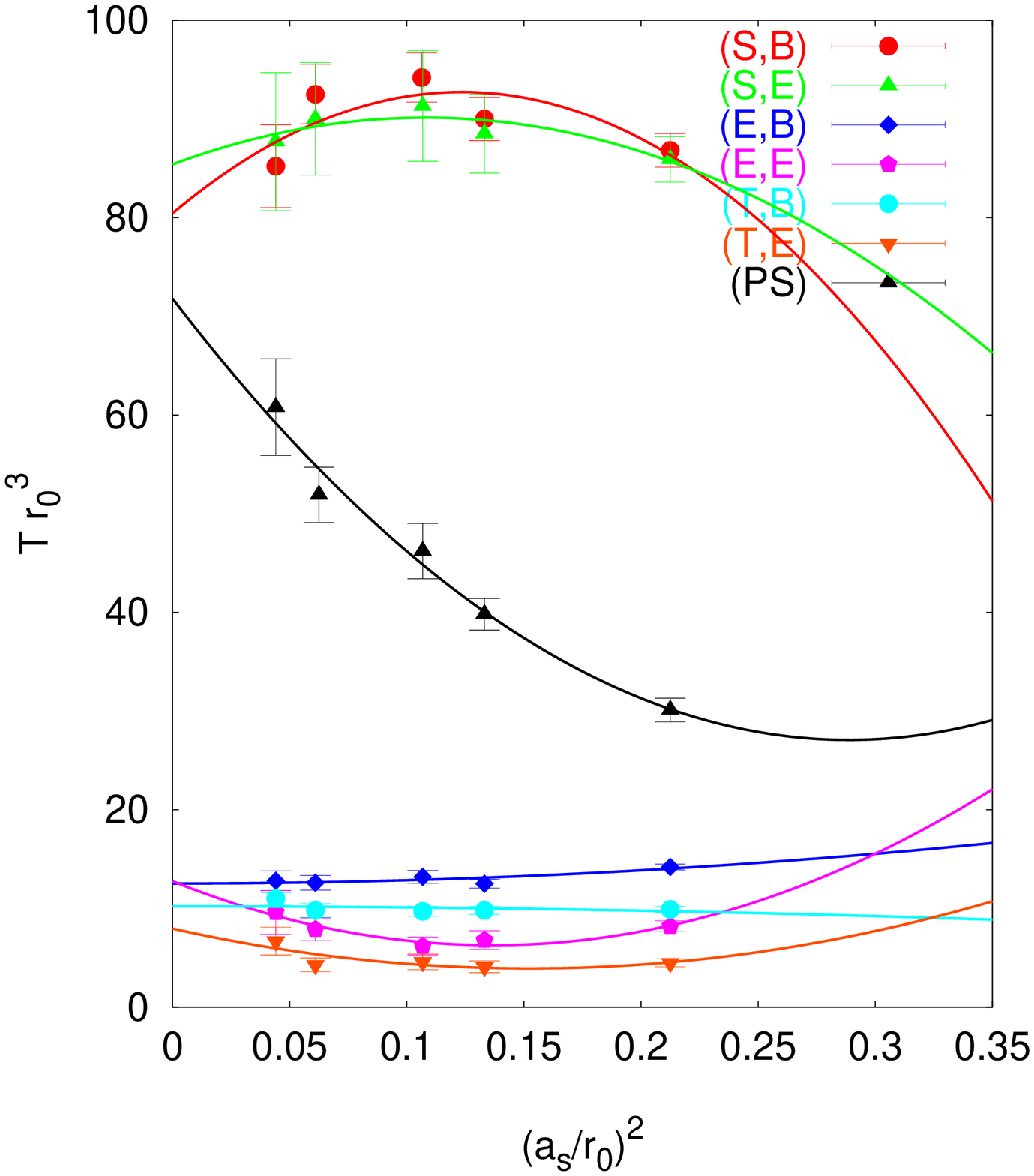}
\includegraphics[height=7.5cm]{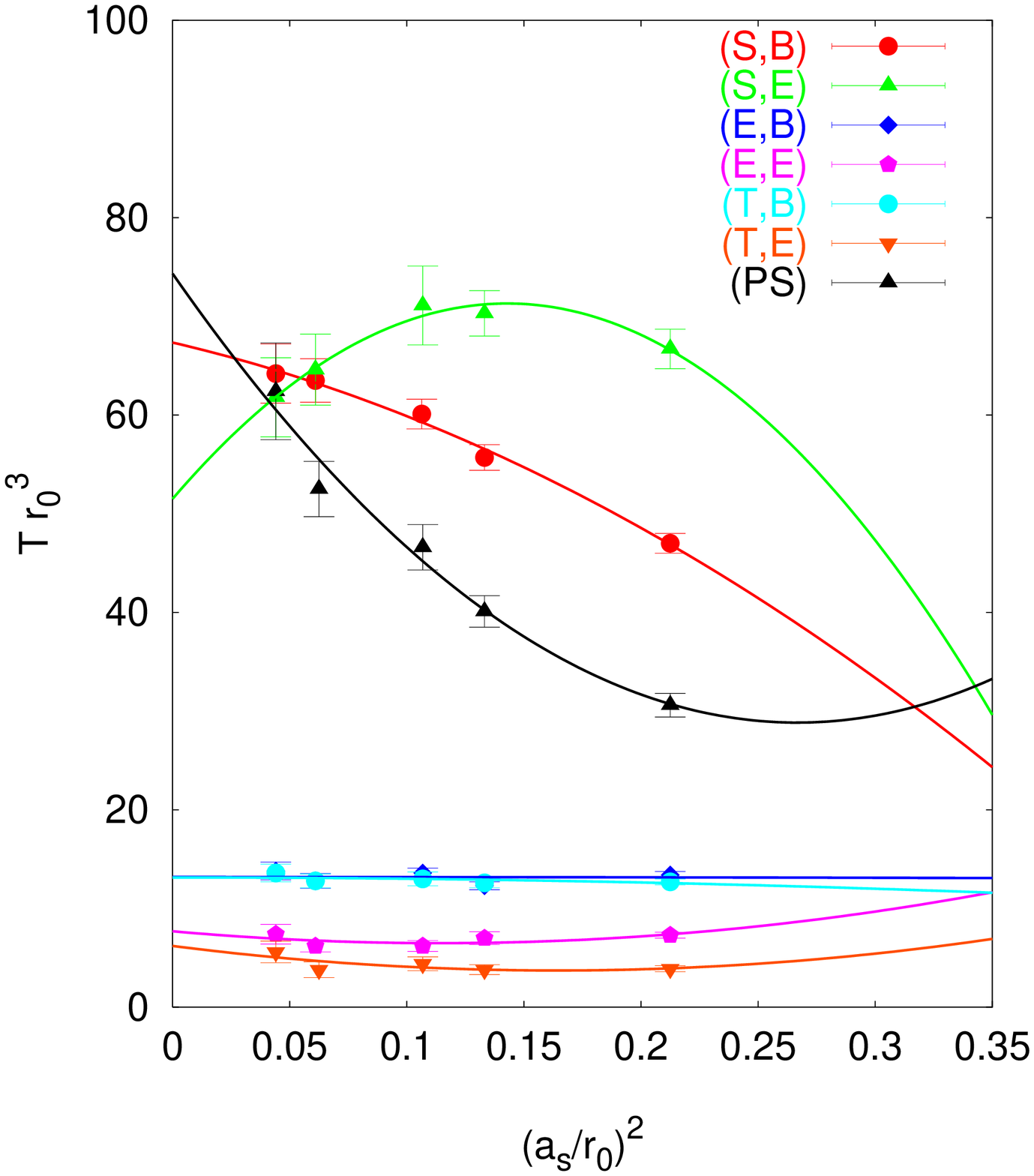}
\caption[]{\label{mtx12} Glueball matrix elements(ME) in terms of
$r_0^3$ against the lattice spacing square $(a_s/r_0)^2$. The
calculated values of ME's are plotted in points, with different
colors corresponding to different channels. The curves are best
fits to the calculated results by the model $T(a_s)r_0^3 =
T(0)r_0^3 + c_2(a_s/r_0)^2 + c_4(a_s/r_0)^4$. The left panel
illustrates the result of Type-I operators, while the right panel
is for Type-II operators. }
\end{figure*}

\par
Based on the discussions above, in the following continuum extrapolation
of matrix elements, we use the results calculated on the lattice
$8^3\times 40$ at $\beta = 2.4$. As for the Type-II operators, we take
the results with the lattice operators defined by
$\hat{F}_{\mu\nu}(x)\hat{F}_{\rho\sigma}(x)$

\par
Using the fitted $M$ and $Y$ parameters listed in
Table~\ref{table24}-\ref{table32}, as well as the physical values
of the lattice spacings $a_s$ listed in Table~\ref{tab:spacing} for each
$\beta$, the lattice matrix
elements $\langle 0|O_L|G\rangle$ can be
obtained by $\langle 0|O_L|G\rangle= \sqrt{2M_G V}Y$. As a result, 
the matrix elements
in units of $r_0^{-3}$ are listed in Table~\ref{phys_me} and illustrated
by Figure~\ref{mtx12}.

\begin{table}
\caption{ \label{phys_me} Final results of matrix elements of lattice
gluonic operators.
The data in table are in the unit of $r_0^{-3}$ with $r_0^{-1}=0.410(20)
\,{\rm MeV}$. }
\begin{center}
\begin{tabular}{c|lllll}\hline \hline
&$\beta=2.4$ & $\beta=2.6$ &$\beta=2.7$  &$\beta=3.0$ & $\beta=3.2$\\
& ($L=8$) & ($L=12$)  &($L=12$) &($L=16$)   &($L=24$) \\ \hline
&&&&&\\
$T(S,B)_I$ &87(2)  & 90(2)   &94(3) &92(3)  &85(4) \\
$T(S,E)_I$ &86(2)  & 89(4)   &91(6) &90(6)  &88(7)\\
&&&&&\\
$T(E,B)_I$ &14.2(3)& 12.5(5) &13.2(7)&12.6(8)   &12.8(10)\\
$T(E,E)_I$ &8.2(6) & 6.8(10) &6.2(9) &7.9(12)   &9.6(22) \\
&&&&&\\
$T(T_2,B)_I$   &9.9(3) & 9.8(4)  &9.7(5) &9.8(7)   &11.0(6)\\
$T(T_2,E)_I$   &4.5(4) & 4.1(6)   &4.6(8) &4.3(7)   &6.7(14)\\
&&&&&\\
$T(PS)_I$   &30(1) &40(2) &46(3) &52(3)  &61(5) \\
\hline
&&&&&\\
$T(S,B)_{II}$ & 47(1) & 56(1) & 60(2) & 63(2) & 64(3) \\
$T(S,E)_{II}$ & 67(2) & 70(2) & 71(4) & 65(4) & 62(4) \\
&&&&&\\

$T(E,B)_{II}$ & 13.4(4) & 12.3(4) & 13.6(5) & 12.8(8) & 13.8(9)\\
$T(E,E)_{II}$ & 7.3(3)  & 7.0(7)  & 6.2(6)  & 6.2(6)  & 7.4(10)\\
&&&&&\\
$T(T_2,B)_{II}$ & 12.7(3) & 12.6(5) & 13.0(7) & 12.8(7) & 13.6(9)\\
$T(T_2,E)_{II}$ & 3.9(3)  & 3.8(5)  & 4.4(5)  & 3.8(6)  & 5.6(10)\\
&&&&&\\
$T(PS)_{II}$   &31(1) &40(2) &47(2) &53(3)  &62(5) \\
\hline
\hline
\end{tabular}
\end{center}
\end{table}

It is clear that the continuum limits can be extrapolated  neither by
the function with
a single $a_s^2$ term nor with a single $a_s^4$ term. We fit them with
the form
\begin{equation}
T(a_s)r_0^3 = T(0)r_0^3 + c_2 \frac{a_s^2}{r_0^2} +
c_4\frac{a_s^4}{r_0^4},
\end{equation}
where $T$ is the final non-renormalized continuum limit results of
the glueball matrix elements.

\par
Note that these matrix elements are all calculated by bare lattice
operators. Before the renormalization of the local operators is
performed, we cannot draw any conclusions of physical interest at
present stage. However, we can give some comments on the different
behaviors of the matrix elements of the Type-I and Type-II
operators from the continuum extrapolation. Based on our
experience from the calculation of glueball masses, the continuum
$SO(3)$ symmetry is approximately restored for all the lattice
spacings we use in this work, since the calculated glueball masses
in $T_2^{++}$ and $E^{++}$ channel are coincident. The left panel
of Fig.~\ref{mtx12} is the plot of the matrix elements of the
Type-I operators, where one can find that the calculated glueball
matrix elements of $T_2$ and $E$ irreps do not show this symmetry
restoration. This is probably due to the definition of the Type-I
operator introduced in Sec. II. The Type-I operators with
different quantum number are defined by the real part of different
Wilson loops composed of different numbers of spatial gauge links,
such that the overall tadpole improvement factors (different
powers of the tadpole parameter $u_s$) are different for $T_2$
and $E$ representation. The conjectured {\it power counting} of
the tadpole parameter $u_s$ for Wilson loops is a naive
approximation and may bring additional deviation to the local
operators. We have carried out the test that the naive tadpole
parameters are replaced by the vacuum expectation value (VEV) of
the corresponding Wilson loops. As we expected, the two matrix
elements tend to coincide better. In contrast with the Type-I
operators, all the Type-II operators are defined by the improved
lattice version of the gauge strength tensor, $\hat{F}_{\mu\nu}$,
and thus have the same correction factor coming from the tadpole
improvement. The right panel of Fig.~\ref{mtx12} shows the
behaviors of the matrix elements of the Type-II operators. It is
clearly seen that the approximate $SO(3)$ symmetry restoration
takes place for $T_2$ and $E$ representation.
\par
The matrix elements of phenomenological interest, denoted $s$, $p$ and
$t$, are defined as
\begin{eqnarray}
(2\pi)^3\delta(\bf{0}) s &=&\langle 0|\int d^3 x
S(x)|G\rangle,\nonumber\\
(2\pi)^3\delta(\bf{0}) p &=&\langle 0|\int d^3 x
P(x)|G\rangle,\nonumber\\
(2\pi)^3\delta(\bf{0}) t\epsilon_{\mu\nu} &=&\langle 0|\int d^3 x
\Theta_{\mu\nu}(x)|G\rangle,
 \end{eqnarray}
where $\epsilon_{\mu\nu}=\epsilon_{\nu\mu}$, $\epsilon_{\mu\mu}=0$,
$\epsilon_{\mu\nu}\epsilon_{\mu\nu}=1$, and $S(x)$, $P(x)$,
$\Theta_{\mu\nu}(x)$ are defined in Eq.~(\ref{lo}) and Eq.~(\ref{def1}).
Recalling that $(2\pi)^3\delta(\bf{0})$ is replaced
by $L^3 a_s^3$ on finite lattices, and combining
Eq.~(\ref{def3}-\ref{def7}), $s$, $p$, and $t$ can be reproduced by the
calculated matrix elements listed in Table~\ref{phys_me} as
\begin{eqnarray}
sr_0^3 &=& 2 (T(S,B)+T(S,E)), \nonumber \\
pr_0^3 &=& 8 T(PS),\nonumber\\
t_E r_0^3 &=& 2| T(E,B)-T(E,E)|,\nonumber \\
t_{T_2}r_0^3 &=& 2|T(T_2,B)-T(T_2,E)|.
\end{eqnarray}

\begin{figure}[htb!]
\includegraphics[height=7.5cm]{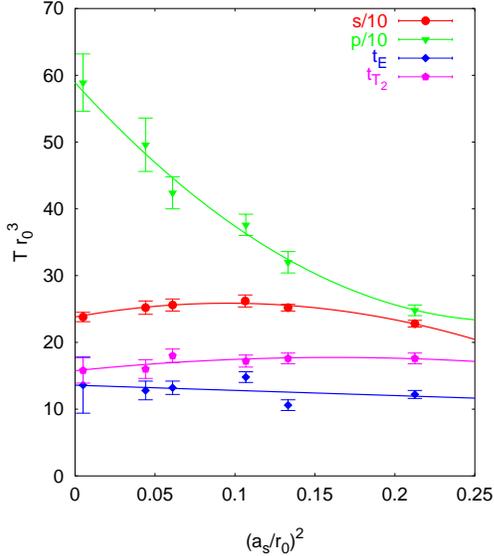}
\caption[]{\label{mtx_cont}
The non-renormalized matrix elements $s$, $p$ and $t$ from type-II 
operators are plotted versus 
lattice spacing. The continuum extrapolations are also carried out.
}
\end{figure}

Fig.~\ref{mtx_cont} and Table~\ref{t_cont} illustrate the behaviors of these matrix
elements with respect to the lattice 
spacings. We only show the data reproduced from the Type-II operators, which can be only 
renormalized in this work (see Section V).   
The continuum limits of $t_{T_2}$ and $t_{E}$ are consistent within
error bars. The deviation of the central values comes mainly from 
the discrepancy of $T(E,E)$ and $T(T_2,E)$.
\begin{table*}
\begin{ruledtabular}
\caption{\label{t_cont} The non-renormalized matrix elements 
$s$, $p$, and $t$ in units of $r_0^{-3}$ are reproduced from the lattice 
results of type-II operators at different lattice spacings. The errors quoted 
here are purely statistical. The continuum limits are also listed for each channel.}
\begin{tabular}{cccccc|c}
 &$\beta=2.4$ & $\beta=2.6$& $\beta=2.7$ & $\beta = 3.0$ & $\beta=3.2$
& Continuum\\
\hline
$sr_0^3$ & 228(5)  & 252(5) & 262(9) & 256(9) &252(10)& 227(7)\\
$pr_0^3$ & 248(8) & 320(16) &376(16)& 424(24)&496(40)& 589(43)\\
$t_E r_0^3$ &12.2(5)&10.6(8) &14.8(8)&13.2(1.0)&12.8(1.4)&13.6(4.1)\\
$t_{T_2} r_0^3$ &
17.6(0.8)&17.6(0.8)&17.2(0.9)&18.0(1.0)&16.0(1.4)&15.8(1.9)
\end{tabular}
\end{ruledtabular}
\end{table*}

\section*{V. The Non-Perturbative Renormalization of Local Gluonic
Operators}
The lattice local gluonic operator $O$ and its continuum
counterpart $O_{cont}$ is related by the renormalization
constant $Z_O(a)$,
\begin{equation}
\label{ren}
O_{cont} = a^{-4} Z_O(a) O.
\end{equation}
The key question is to choose a proper normalization condition, so that
the renormalization constant can be determined. In this work, we choose
the energy-momentum tensor in the glueball state
as the normalization condition.

The energy-momentum tensor for the pure gauge theory
\begin{equation}
\bar{T}_{\mu\nu} = \bar{T}_{\nu\mu} = -\frac{1}{2} \delta_{\mu\nu} {\rm 
Tr}F^2 +
2{\rm Tr}F_{\mu\alpha}F_{\nu\alpha}
\end{equation}
satisfies $\partial_{\mu} \bar{T}_{\mu\nu} = 0$ and does not need an 
overall renormalization in the continuum. More specifically,
$\bar{T}_{00} = Tr({\bf E}^2-{\bf B}^2)$). At the classical level, 
$\bar{T}_{\mu\nu}$ is traceless. When quantum corrections are included,
the renormalized energy-momentum tensor, $T_{\mu\nu}$, takes a 
non-vanishing trace part, $\hat{T}_{\mu\nu}$, which comes from the 
anomalous breaking of scale invariance and is called the QCD trace 
anomaly, 
 \begin{equation}
\hat{T}_{\mu\nu} = -\frac{1}{4}\frac{\beta(g)}{2g} \delta_{\mu\nu}F^2
= \frac{11}{32\pi^2}\delta_{\mu\nu}Tr(g^2{\bf B}^2 + g^2{\bf E}^2),
\end{equation}
where $\beta(g)=-\beta_0 g^3/(4\pi)^2$ is the $\beta$ function of QCD
to the lowest order of the coupling constant $g$ with $\beta_0 = 11$ in 
the pure gauge case. Thus, $T_{\mu\nu}$ can be 
written explicitly as the sum of the traceless and trace parts,
\begin{equation}
T_{\mu\nu}=\bar{T}_{\mu\nu}+\hat{T}_{\mu\nu}.
\end{equation}

On the other hand, $T_{\mu\nu}$ defines the Hamiltonian operator of the 
theory 
\begin{equation}
H = \int d^3 x T_{00}({\bf x},0),
\end{equation}
which is finite and scale independent. In the rest frame of a glueball, 
the matrix element of the Hamiltonian $H$ in the glueball state is 
just the glueball's rest mass,
\begin{equation}
\label{m}
M_G = \frac{\langle G(M_G,{\bf p}=0)|\int d^3 x T_{00}({\bf
x},0)|G(M_G,{\bf p}=0)\rangle}
            {\langle G|G \rangle}.
\end{equation}
If the glueball states $|G(E,{\bf p})\rangle$ are normalized as
\begin{equation}
\langle G(E_p,{\bf p})|G(E_p,{\bf p'})\rangle = 2
E_p(2\pi)^3\delta^3 ({\bf p}-{\bf p'}),
\end{equation}
we have $\langle G(M_G,{\bf p}=0)|\int d^3 x T_{00}({\bf
x},0)|G(M_G,{\bf p}=0)\rangle = 2 M_G^2 (2\pi)^3\delta^3({\bf
0})\equiv 2 
p^0p^0(2\pi)^3\delta^3({\bf
0})$, which implies that
\begin{equation}
\label{tmn}
\langle G(p)| \int d^3 x T_{\mu\nu}({\bf x},0) |G (p)\rangle = 2
p^{\mu}p^{\nu}(2\pi)^3\delta^3({\bf 0}).
\end{equation}
Combining Eq.~(\ref{m}) and~(\ref{tmn}),
we get the normalization conditions in the glueball's rest frame,
\begin{widetext} 
\begin{eqnarray}
\label{norm}
\langle G(p)|\int d^3 x\bar{T}_{\mu\nu}({\bf x},0)|G(p)\rangle 
&=&
2(p^{\mu}p^{\nu}-\frac{1}{4} M_G^2 \delta_{\mu\nu})(2\pi)^3
\delta^3({\bf 0}),
\nonumber\\
\langle G(p)|\int d^3 x\hat{T}_{\mu\nu}({\bf x},0)|G(p)\rangle &=& 
\frac{1}{2} M_G^2
\delta_{\mu\nu} (2\pi)^3 \delta^3({\bf 0}).\nonumber\\
\end{eqnarray}
\end{widetext}
\par
According to the definitions in Eq.~\ref{def5} and
Eq.~\ref{def6}, the $E$ and $T_2$ operators are related to $\bar{T}_{ij}$ by 
\begin{eqnarray}
O_1^{(E,E)}-O_1^{(E,B)}&\propto& \bar{T}_{11}-{\bar T}_{22},\nonumber \\
O_2^{(E,E)}-O_2^{(E,B)}&\propto& 2{\bar T}_{33}-\bar{T}_{11}-\bar{T}_{22}, \nonumber\\
O_i^{(T_2,E)}-O_i^{(T_2,B)}&\propto& \epsilon_{ijk}\bar{T}_{jk},
\end{eqnarray} 
whose matrix elements in a zero-momentum glueball state vanish, thus we 
cannot renormalize $E^{++}$ and $T_2^{++}$ operators directly by the normalization 
condition in Eq.~\ref{norm} in the glueball rest frame. 
A possible way around this difficulty is to assume the almost 
restoration of Lorentz invariance at the lattice spacings we use in this work, 
so the overall renormalization constant of the lattice version of $\bar{T}_{\mu\nu}$ 
can be determined by one of its components. 
In fact, this is justified to some extent by two
facts. First, it is argued that the rotational invariance can be
restored if the scale parameter $z=m_{A_1^{++}}L$ is larger than 
10~\cite{lusher},
and our smallest lattice gives the value $z = 0.31\times 5\times 8 =
12.4 $ which meets this requirement. Secondly, in our lattice
calculations of the mass spectrum and matrix elements, the coincidence of
$E^{++}$ channel and $T_2^{++}$ channel implies that this rotational
restoration is actually realized. Based on the discussion above, we 
choose the component $\bar{T}_{00}$ to do 
the renormalization of the tensor
operator in this work. In the practical study, we calculate 
the matrix elements $\langle G|O_{\pm}|G\rangle_{\rm lat}$, 
where $O_{\pm}$ is the operator $g^2 Tr({\bf E}^2\pm {\bf B}^2)$ and 
$|G\rangle = |A_1^{++}\rangle$, $|E^{++}\rangle$, or $|T_2^{++}\rangle$. 
As we have addressed before, $O_{-}$ is proportional to $\bar{T}_{00}$, 
and $O_{+}$ is proportional to the trace anomaly $\hat{T}_{00}$.
Thus, the renormalization constants of the scalar and tensor operators 
can be extracted from these matrix elements. 
\par
The matrix 
elements $\langle G|O_{\pm}|G\rangle_{\rm lat}$ can be obtained by
calculating the three-point function
\begin{widetext}
\begin{eqnarray}
C_3(t_1, t_2) &=& \langle 0|O_s(-t_1)O_{\pm}(0)O_s(t_2)|0\rangle =
\sum\limits_{m,n}\frac{1}{4E_n E_m V^2}\langle 0|O_S(-t_1)|n\rangle
                                 \langle n|O_{\pm}(   0)|m\rangle
                                 \langle m|O_S( t_2)|0\rangle\nonumber\\
         &=& \sum\limits_{m,n}\frac{1}{4E_n E_m V^2}
        \langle 0|O_S(   0)|n\rangle \langle n|O_{\pm}(   0)|m\rangle
        \langle m|O_S(   0)|0\rangle e^{-E_n t_1} e^{-E_m t_2}\nonumber\\
         &\rightarrow& \frac{1}{(2M_G V)^2}
                        \langle 0|O_s(0)|G\rangle |^2
                        \langle G|O_{\pm}(0)|G\rangle
e^{-M_G(t_1+t_2)}~~(t_1,~t_2\rightarrow \infty)
\end{eqnarray}
\end{widetext}
where $O_S(t)$ is the smeared zero-momentum operator which generates the 
ground state $|G\rangle$ from the vacuum, and $M_G$ the ground state 
mass. Using the
asymptotic form of the two-point function
\begin{equation}
C_{SS}(t) \sim \frac{1}{2M_G V}|\langle 0|O_S(0)|G\rangle|^2
                             e^{-M_G t}~~(t \rightarrow \infty)
\end{equation}

\noindent and dividing the three-point function by proper two-point 
function, one gets
\begin{equation}
\frac{C_3(t_1,t_2)}{C_{SS}(t_1)} \approx \frac{1}{2M_G V}\langle
G|O_{\pm}(0)|G\rangle  e^{-M_G t_2}~~(t_1, t_2 \rightarrow \infty).
\end{equation}
In the practical data processing, by analogy with the extraction of the 
matrix element, we fit $C_2(t)$ and $C_3(t,t_0)$ simultaneously
with the fitting model 
\begin{eqnarray}
C_{SS}(t)&=&X^2 e^{-Mt}  \nonumber \\
C_3(t,t_0) &=& X^2 W e^{-M(t+t_0)},
\end{eqnarray}
where $t_0=1$. $W$ is related to the matrix element $ \langle
G|O_{\pm}(0)|G\rangle$ by 
\begin{equation}
\langle G|O_{\pm}(0)|G\rangle = 2 M_G V W.
\end{equation}
From Eq.~(\ref{ren}), 
Eq.~(\ref{norm}) and the fitted
matrix elements $\langle G|O_{\pm}|G\rangle_{\rm lat}$, the 
renormalization
constants for scalar and tensor gluonic operators can be derived as
\begin{eqnarray}
\frac{1}{Z_S(a_s)} &=& \frac{11}{32\pi^2}\frac{2}{M_G^2 a_s^4}
 \langle G|O_{+}|G\rangle_{\rm lat},\nonumber\\
\frac{1}{Z_T(a_s)} &=& \frac{1}{g^2}\frac{2}{3M_G^2 a_s^4}
 \langle G|O_{-}|G\rangle_{\rm lat},
\end{eqnarray}
where the coupling constant $g^2$ comes from the relation $\bar{
T}_{00}=(1/g^2)O_{-}$.
\par
Unfortunately, the gluonic three-point function is far more noisy than
the gluonic two-point function in Monte Carlo calculation. In this work,
the renormalization of gluonic operators is performed only at $\beta=2.4$ on
the lattice $8^3\times 40$ with $\xi=5$. As many as 100,000 measurements
are carried out, the signals of three point functions are still
weak with large fluctuation. In the practical computation, the matrix 
elements of Type-I and Type-II operators of $O^{(S,B)}$ and $O^{(S,E)}$ are 
all calculated. It is found that
the three-point functions involving Type-I operators are so noisy that
the matrix elements cannot be extracted reliably. The three-point
functions involving Type-II operators behave better, from which we 
obtain the renormalization constants. The
matrix elements $ \langle
G|O_{\pm}|G\rangle_{\rm lat}$ and the resultant renormalization 
constants at
$\beta = 2.4 $ are listed in Table~\ref{const}.

\begin{table}
\begin{ruledtabular}
\caption{\label{const} The matrix elements of operator $O_{\pm}$ (Type-II) in 
glueball states are listed. $S$ represents the scalar glueball 
state, while $T(E)$ and $T(T_2)$ denotes the tensor glueballs in $E$ and 
$T_2$ irreps, respectively. The renormalization constants extracted from 
these matrix elements are given also. } 
\begin{tabular}{c|rr|rr}
$|G\rangle$ & $\langle G|O_{+}|G\rangle_{\rm lat}$ & $Z_S $ &
$\langle G|O_{-}|G\rangle_{\rm lat}$ & $Z_T $\\
\hline
$S$ &  32(9)   & 1.1(3) & 13(5)  &   0.7(3) \\
$T(E)$ & 102(16)  & 1.0(2) & 51(15) & 0.53(15) \\
$T(T_2)$ & 101(16)  & 1.0(2) & 53(15) & 0.51(15)
\end{tabular}
\end{ruledtabular}
\end{table}

\begin{table*}
\begin{ruledtabular}
\caption{\label{top} 
The non-renormalized topological susceptibility $\chi_L(a_s)$ calculated 
at different lattice spacing are listed in physical units. Also listed 
are the ratios of $\chi_L^{1/2}(a_s)$ and the non-renormalized matrix elements
of pseudoscalar $p(a_s)$ after rescaled by their continuum extrapolation values
$\chi_L^{1/2}(0)$ and $p(0)$. }
\begin{tabular}{c|cccccc}
$\beta$         &  2.4   & 2.6    & 2.7  & 3.0   & 3.2  & continuum\\
\hline
$\chi_L^{1/4}(a_s)({\rm MeV})$ &  242(4)& 277(5) &299(5)&323(6)& 351(8)& 
391(15)\\
$p(a_s) r_0^3$	&  248(8)&320(16) &376(16)&424(24)&496(40)&589(43)\\
$\frac{p(a_s) \chi_L^{1/2}(0)}{\chi_L^{1/2}(a_s) p(0)}$
		& 1.10(7)   &1.08(9) & 1.09(8) & 1.06(10) & 1.05(11) & 1.00(13)\\
\end{tabular}
\end{ruledtabular}
\end{table*}
\begin{figure}[htb!]
\includegraphics[height=7.5cm]{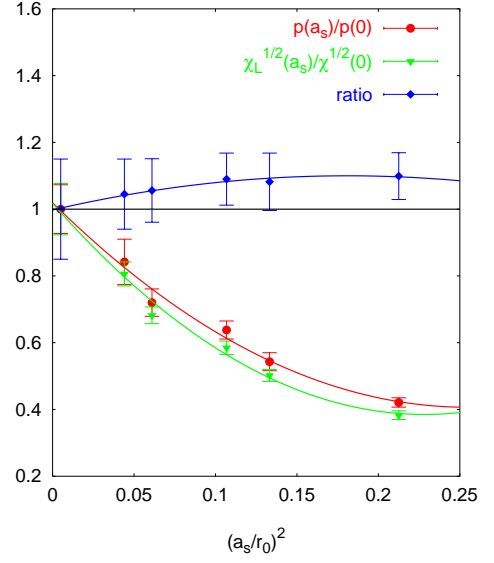}
\caption[]{\label{chips}
The $a_s$-dependences of $\chi_L^{1/2}(a_s)$ and $p(a_s)$ are plotted.
For comparison, $\chi_L^{1/2}(a_s)$ and $p(a_s)$ are rescaled by 
their continuum-extrapolated values, say, $\chi_L^{1/2}(0)$ and $p(0)$,
respectively. One can find that the $a_s$-dependences of
$\chi_L^{1/2}(a_s)$ and $pr_0^3(a_s)$ are very similar and their ratios are 
fairly constant.
}
\end{figure}

\par
As for the renormalization of the pseudoscalar operator, we tentatively 
use the phenomenological value of the topological susceptibility, 
$\chi^{1/4}=180\,{\rm MeV}$, as the normalization condition. The 
quantity $\chi$ 
is defined by 
\begin{equation}
\chi = \int d^4 x \langle q(x)q(0)\rangle,
\end{equation}
where the topological charge density $q(x)$ is proportional to the 
pseudoscalar operator $P(x)$ (defined in Eq.~(\ref{lo}) as $ q(x) = 
\frac{1}{32\pi^2}P(x)$. The lattice version of $\chi$, denoted by 
$\chi_L$, is defined by the lattice pseudoscalar operator $O(PS)(x)$ as
\begin{equation}
\chi_L  = \frac{\xi}{(32\pi^2)^2 L^3 Ta_s^4}\sum\limits_{x,y} 
\langle O^{(PS)}(x)O^{(PS)}(y)\rangle,
\end{equation}
where $L$ and $T$ are the lattice sizes in spatial and temporal 
directions, respectively. 
\par
We have calculated $\chi_L(a_s)$ for all the five $\beta$, at each 
$\beta$ 20000 measurements are carried out. In Table~\ref{top} are 
listed the results of $\chi_L^{1/4}(a_s)$ in units of MeV for different $\beta$, 
as well as the non-renormalized continuum value $\chi_L^{1/4}(0)$ after the 
continuum extrapolation $a_s\rightarrow 0$. It is obvious that the value of  
$\chi_L^{1/4}(a_s)$ increases along with the decreasing of lattice spacing. 
For comparison, the non-renormalized matrix elements $pr_0^3(a_s)$ at different $\beta$'s
are also listed in Table~\ref{top}. In Fig.~\ref{chips} are plotted the $a_s$-dependences of 
$\chi_L^{1/2}(a_s)$ and $pr_0^3(a_s)$ (rescaled by their continuum-extrapolated values, 
respectively), as well as their ratios, where one can find that the $a_s$-dependences of
$\chi_L^{1/2}(a_s)$ and $pr_0^3(a_s)$ are very similar and their ratios are fairly constant.
Using $\chi^{1/4}=180\,{\rm MeV}$, the renormalization constant $Z_P$ of pseudoscalar 
operator in the continuum limit can be extracted as
\begin{equation}
Z_P= \frac{\chi^{1/2}}{\chi_L^{1/2}(0)}\approx 0.21(2)
\end{equation}

\section*{VI. RESULTS AND DISCUSSION}

After the continuum extrapolation and the renormalization of the local
operators, we can discuss the physical implication of our lattice
results.
\par
As we addressed above, we have tried to extract the renormalization 
constants for scalar and tensor operator only at $\beta=2.4$. In fact,
the signals of the three-point functions of type-I operator are very 
poor, so the renormalization constants $Z_S$ and $Z_T$ are obtained only 
for Type-II operators. We are lucky with this situation because all the 
Type-II operators are made up of the lattice version of the gauge field 
strength, $\hat{F}_{\mu\nu}$, thus they all have the same normalization 
constant, say, the same tadpole-improvement factor. Therefore, for 
Type-II operator, the $Z_T$ extracted from $\bar{T}_{00}$ can be applied to 
other components involved in the glueball matrix elements calculated in 
this work.  
\par
Considering Eq.~(\ref{def1}) and~(\ref{def2}), the 
non-renormalized matrix elements of the $S(x)$, $P(x)$, and $\Theta(x)$ 
in the continuum can be obtained by the
lattice results, and are shown in Table~\ref{t_cont}. We notice that, 
although the $(S, B)$ and $(S, E)$ matrix elements have sizeable 
$a_s$ dependence for the Type-II operator, as seen in Fig.~\ref{mtx12}, 
the total scalar matrix element, which is the sum of the two, is much 
flatter in $(a_s/r_0)^2$. This is also the case for the tensor matrix 
element. With the observation that the non-renormalized matrix elements 
of the scalar and the tensor depend on the lattice spacing $(a_s/r_0)^2$ very 
mildly, we speculate that there is not a large lattice 
spacing dependence in the renormalization constant, and will use 
$Z_S$ and $Z_T$ computed at $\beta = 2.4$ as an approximation 
of the renormalization constants in the continuum limit. We shall check this in 
the future when computer resources are available 
for high statistics calculation at larger $\beta$. 

\par
Taking the average of the renormalization constant $Z_S$ from
Table~\ref{const}, we get a continuum extrapolated value for the matrix
element $\langle 0|S|0^{++}\rangle =  15.6 \pm 3.2\,{\rm GeV}^3$. 
Based on the scaling properties of QCD and trace anomaly, both the QCD 
sum
rule~\cite{sumrule} and the soft meson theorem~\cite{softmeson} lead to
an estimate that relates the scalar glueball matrix element to the
gluon condensate,
\begin{equation}
\label{sum1}
\langle 0|S|0^{++}\rangle = 16\pi^2\sqrt{\frac{G_0}{2b}}M_G,
\end{equation}
where $G_0=\langle 0|\frac{\alpha_s}{\pi}G_{\mu\nu}^a
G_{\mu\nu}^a|0\rangle$ is the gluon condensate, $b=(11/3)N_c-(2/3)N_f$
and $M_G$ the scalar glueball mass. Taking $N_f=0$, $G_0=0.012\,{\rm 
GeV}^4$,
and $M_G\approx 1.7\,{\rm GeV}$, this matrix element is estimated to be 
$\sim
6.3$ GeV, which is about two and a half times smaller than our lattice 
result. This discrepancy might be attributable to the fact that the 
quenched lattice calculation gives a gluon condensate which is about 
$0.14\pm 0.02 {\rm GeV}^4$~\cite{giacomo}. This is larger by an order of 
magnitude than that used in QCD sum rule. If the relation 
Eq.~(\ref{sum1}) 
still holds in the pure gauge theory, using the quenched gluon
condensate, the estimated scalar matrix element is estimate to be  
$(21\pm 1){\rm GeV}^3$ which is in good agreement with our quenched 
lattice calculation.
\par
For the pseudoscalar, with the renormalization constant $Z_P=0.21(2)$ determined 
in last section,
the lattice calculation gives the result $\langle 0|P|0^{-+}\rangle 
\approx 8.6\pm 1.3\,{\rm 
GeV}^3$. It has been 
proposed that there is an approximate chiral symmetry between the scalar 
and pseudoscalar glueballs~\cite{chiral}. A sum rule is derived from an
effective action which relates the topological susceptibility $\chi$ in
the pure gauge case to the gluon condensate $G_0$,
\begin{equation}
\chi = \eta^{-2} \sqrt{\frac{G_0}{2b}},
\end{equation}
where $\eta\approx 0.7$.
Using our lattice results, the degree of chiral symmetry $\eta$ can be
obtained from the ratio of the pseudoscalar to scalar matrix
elements~\cite{chiral},
$\eta_L = \langle 0|P|0^{-+}\rangle/(2\langle 0|S|0^{++}\rangle)\approx
0.3$, which is also more than two times smaller than the result of 
QCD sum rule. These facts hint that there may be a substantial quenching 
effect in the matrix element of the scalar.
\par
We can also estimate the glueball contribution to the topological
susceptibility by the lattice matrix element and glueball mass,
\begin{equation}
\chi_{\rm glueball} = \left(\frac{\langle 0|P|0^{-+}\rangle}{32\pi^2
M_{0^{-+}}}\right)^2\approx (103 \,{\rm MeV})^4\sim 0.11 \chi,
\end{equation}
which implies that the pseudoscalar glueball gives an 
appreciable 11\% contribution to the topological susceptibility.
\par
In the tensor channel, the glueball matrix element is extrapolated 
to $1.0\pm 0.2\,{\rm GeV}^3$ in the continuum, which is the average of 
results of $E$ and $T_2$ channels. In the calculation, it is found 
that in the lattice spacing
range we use, the glueball mass and matrix elements are approximately
independent of the lattice spacing, this implies that the lattice
artifacts might be neglected here.
If the renormalization constant
$Z_T\approx 0.52(15)$ of the tensor operator does not change much in 
the range of lattice spacing and applies to all the $\beta$ 
values in this work, the
renormalized matrix element of tensor operator is $0.52\pm 0.19\,{\rm 
GeV}^3$, which is in agreement with the prediction $ 0.35\,{\rm GeV}^3$ 
from the tensor dominance model~\cite{tensor} and QCD sum 
rule~\cite{sum3} for the tensor mass around $2.2 \,{\rm GeV}$.

\begin{table}
\begin{ruledtabular}
\caption{\label{summary1}  The final glueball spectrum in physical 
units. In column 2, the first error is the statistical uncertainty 
coming from the continuum extrapolation, the second one is the 1\% 
uncertainty resulting from the approximate anisotropy. In column 3, 
the first error comes from the combined uncertainty of $r_0 M_G$, the 
second from the uncertainty of $ r_0^{-1}=410(20)\,{\rm MeV}$ 
} \begin{tabular}{ccc}
$J^{PC}$& $r_0 M_G$ &	$M_G\,({\rm MeV}$) \\
\hline
$0^{++}$ & 4.16(11)(4)   &    1710(50)(80)     \\
$2^{++}$ & 5.83(5)(6)    &    2390(30)(120)    \\
$0^{-+}$ & 6.25(6)(6)    &    2560(35)(120)    \\
$1^{+-}$ & 7.27(4)(7)    &    2980(30)(140)     \\
$2^{-+}$ & 7.42(7)(7)    &    3040(40)(150)     \\
$3^{+-}$ & 8.79(3)(9)    &    3600(40)(170)     \\
$3^{++}$ & 8.94(6)(9)    &    3670(50)(180)     \\
$1^{--}$ & 9.34(4)(9)    &    3830(40)(190)     \\
$2^{--}$ & 9.77(4)(10)   &    4010(45)(200)     \\
$3^{--}$ & 10.25(4))(10) &    4200(45)(200)     \\
$2^{+-}$ & 10.32(7)(10)  &    4230(50)(200)     \\
$0^{+-}$ & 11.66(7)(12)  &    4780(60)(230)     
\end{tabular}
\end{ruledtabular}
\end{table}

\section*{VII. Conclusion}
The glueball mass spectrum and glueball-to-vacuum matrix elements are 
calculated on anisotropic lattices in this work. The calculations are 
carried out at five lattice spacings $a_s$'s which range from 
$0.22\,{\rm fm}$ to $0.10\,{\rm fm}$. Due to the implementation of the 
improved gauge action and improved gluonic local operators, the lattice 
artifacts are highly reduced. The finite volume effects are carefully 
studied with the result that they can be neglected on the lattices we 
used in this work. 
\par
\begin{figure}[t!]
\includegraphics[height=7.5cm]{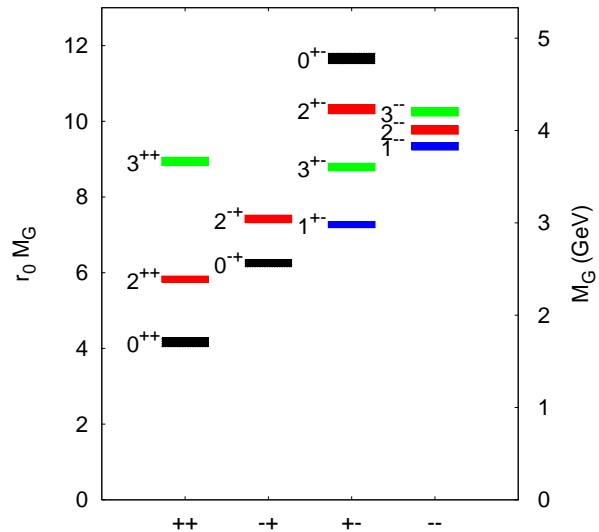}
\label{continuum}
\caption[]{\label{summary2}
The mass spectrum of glueballs in the pure $SU(3)$ gauge theory.
The masses are given both in terms of $r_0$ ($r_0^{-1}=410\,{\rm MeV}$) 
and in ${\rm GeV}$. The height of each colored box indicates the 
statistical uncertainty of the mass.
}
\end{figure}

As to the glueball spectrum, we have carried out calculations 
similar to the previous work~\cite{old3} on much larger and finer 
lattices, so that the liability of the continuum limit extrapolation
is reinforced. Our results of the glueball spectrum is summarized in 
Tab.~\ref{summary1} and Fig.~\ref{summary2}. 

\par
After the non-perturbative renormalization 
of the local gluonic operators, we finally get the matrix elements of 
scalar($s$), pseudoscalar($p$), and tensor operator ($t$) with the results
\begin{eqnarray}
s &=& 15.6\pm 3.2~({\rm GeV})^3 \nonumber \\
p &=& 8.6 \pm 1.3~({\rm GeV})^3 \nonumber \\
t &=& 0.52\pm 0.19~({\rm GeV})^3,
\end{eqnarray}
where the errors of $s$ and $t$ come mainly from the errors 
of the renormalization constants $Z_S$ and $Z_T$. The more precise
calculation of $Z_S$ and $Z_T$ will be carried out in later work.
 
\begin{acknowledgments}
This work is supported in part by U.S. Department of Energy under
grants DE-FG05-84ER40154 and DE-FG02-95ER40907. The computing
resources at NERSC (operated by DOE under DE-AC03-76SF00098) and
SCCAS (Deepcomp 6800) are also acknowledged. Y. Chen is partly supported by NSFC
(No.10075051, 10235040) and CAS (KJCX2-SW-N02). C. Morningstar is also 
supported by NSF under grant PHY-0354982.
\end{acknowledgments}



\begin{thebibliography}{99}

\bibitem{berg} B. Berg and A. Billoire, Nucl. Phys. {\bf B221}, 109
(1983).
\bibitem{old1}
 G. Bali, {\it et al}. (UKQCD Collaboration), Phys. Lett. {\bf B 309},
378 (1993).

\bibitem{old2} C. Michael and M. Teper, Nucl. Phys. {\bf B314},
347 (1989).

\bibitem{old3} C. Morningstar and M. Peardon, Phys. Rev. {\bf D56}, 3043
(1997); Phys. Rev. {\bf D60}, 034509 (1999).

\bibitem{lepage} G.P. Lepage and P.B. Mackenzie, Phys. Rev. {\bf D48},
2250 (1993).

\bibitem{sommer94} R. Sommer, Nucl. Phys. {\bf B411}, 839 (1994).

\bibitem{liu1} K.F. Liu, B.A. Li, and K. Ishikawa, Phys. Rev. {\bf
D40}, 3648 (1989).

\bibitem{liu2} Y. Liang, K.F. Liu, B.A. Li, S.J. Dong, and K. Ishikawa,
Phys. Lett. {\bf B307}, 375 (1993).

\bibitem{jixd} X. Ji, Phys. Rev. Lett. {\bf 74}, 1071 (1995).
\bibitem{lusher} M. L\"{u}scher, Phys. Lett. {\bf 118B}, 387 (1982).
\bibitem{dong} S.J. Dong, {\it et al.}, Nucl. Phys. Proc. Suppl. {\bf
63}, 254 (1998).

\bibitem{stokes}Yu.A. Simonov, Phys. Atom. Nucl. {\bf 50}, 213, (1989).
\bibitem{sumrule}V.A. Novikov, M.A. Shifman, A.I. Vainshtein, and
Zakharov, Nucl. Phys. {\bf B165}, 67 (1980).
\bibitem{softmeson} J. Ellis and J. Lanik, Phys. Lett. {\bf 150B}, 289 
(1985).
\bibitem{giacomo} A.Di. Giacomo, H.G. Dosch, V.I.
Shevchenko, Yu.A. Simonov, Phys. Rept. {\bf 372}, 319
(2002), hep-lat/0007223.

\bibitem{chiral} J.M. Cornwall and A. Soni, Phys. Rev. {\bf D29} 1424
(1984).

\bibitem{tensor} K. Ishikawa, I. Tanaka, K.F. Liu, and B.A. Li, Phys.
Rev. {\bf D37}, 3216 (1988).

\bibitem{sum3} S. Narison, Z. Phys. {\bf C26}, 209 (1984).

\end{thebibliography}
\end{document}